\documentclass[12pt,draftclsnofoot,onecolumn,journal]{IEEEtran} 

\usepackage{graphicx}
\usepackage{epstopdf}
\usepackage{float}
\usepackage{algorithm,algorithmic}
\usepackage{array}
\usepackage{amsmath}
\usepackage{amssymb}
\usepackage{mdwmath}
\usepackage{mdwtab}
\usepackage{eqparbox}

\usepackage{fixltx2e}
\usepackage{cases}
\usepackage{bm}
\usepackage{multirow}
\usepackage{psfrag}
\usepackage[usenames]{color}
\usepackage{mathtools}

\usepackage{fixmath}
\usepackage{mathdots}
\usepackage{latexsym}
\usepackage{amsmath,amssymb}

\usepackage{bm}
\usepackage{footmisc}

\usepackage[T1]{fontenc}
\usepackage{url}
\usepackage{cite}
\newcommand{\subparagraph}{}
\usepackage{titlesec}

\newtheorem{theorem}{\bf Theorem}
\newtheorem{proposition}{Proposition}
\newtheorem{lemma}{\bf Lemma}

\newcommand{\diag}{\mathop{\mathrm{diag}}}
\newcommand{\tr}{\mathop{\mathrm{Tr}}}

\begin{document}

\title{Generalized Beamspace Modulation Using Multiplexing: A Breakthrough in mmWave MIMO}
\author{Shuaishuai~Guo,~\IEEEmembership{Member, IEEE,}
        Haixia~Zhang,~\IEEEmembership{Senior Member, IEEE,}\\
        Peng Zhang,~\IEEEmembership{Member, IEEE,}
Pengjie Zhao, Leiyu Wang,  \\and   Mohamed-Slim Alouini,~\IEEEmembership{Fellow, IEEE}
\thanks{S. Guo and M. -S. Alouini are with King Abdullah University of Science and Technology, Thuwal, Kingdom of Saudi Arabia, Saudi Arabia, 23955 (email: \{shuaishuai.guo; slim.alouini\}@kaust.edu.sa).}
\thanks{H. Zhang, P. Zhao  and L. Wang  are with Shandong University, Jinan, China, 250061 (e-mail: haixia.zhang@sdu.edu.cn; \{pengjie.zhao; leiyu\textunderscore wang\}@mail.sdu.edu.cn).}
\thanks{P. Zhang is with the School of Computer Engineering, Weifang University, Weifang 261061, China (e-mail: sduzhangp@163.com).}
\thanks{Manuscript received Oct. XX, 2018}}
\markboth{Submitted to IEEE JOURNAL ON SELECTED AREAS IN COMMUNICATIONS}%
{Guo \MakeLowercase{\textit{et al.}}:Generalized Beamspace Modulation Using Multiplexing: A Breakthrough in mmWave MIMO}
\maketitle
\newpage
\begin{abstract} 
Spatial multiplexing (SMX) multiple-input multiple-output (MIMO)  over the best beamspace was considered as the best solution  for millimeter wave (mmWave) communications regarding spectral efficiency (SE), referred as the best beamspace selection (BBS) solution. The equivalent MIMO  water-filling (WF-MIMO) channel capacity was treated as an unsurpassed SE upper bound. Recently, researchers have proposed various schemes trying to approach the benchmark and the performance bound. But, are they the real limit of mmWave MIMO systems with reduced radio-frequency (RF) chains? In this paper, we challenge the benchmark and the corresponding bound by proposing a better transmission scheme that achieves higher SE, namely the Generalized Beamspace Modulation using Multiplexing (GBMM). Inspired by the concept of spatial modulation, besides the selected beamspace, the selection operation is used  to carry information. We prove that GBMM is superior to BBS in terms of SE and can break through the well known ``upper bound''. That is,  GBMM renews the upper bound of the SE. We investigate SE-oriented precoder  activation probability optimization, fully-digital precoder design, optimal power allocation and hybrid precoder design for GBMM. A gradient ascent algorithm is developed to find the optimal solution, which is applicable in all signal-to-noise-ratio (SNR) regimes. The best solution is derived in the high SNR regime. Additionally, we investigate the hybrid receiver design and deduce the minimum number of receive RF chains configured to gain from GBMM in achievable SE. We propose a coding approach to realize the optimized precoder activation. An extension to mmWave broadband communications is also discussed. Comparisons with the benchmark (i.e., WF-MIMO channel capacity) are made under different system configurations to show the superiority of GBMM.
\end{abstract}

\newpage
\IEEEpeerreviewmaketitle

\section{Introduction}
\IEEEPARstart{M}{illimeter} wave (mmWave) communication as a promising technology in the fifth-generation (5G) wireless networks has received much attention in recent years \cite{Bai2014}. It has a large amount of available spectrum resource, which enables 5G to meet the increasing data traffic demand. However, its coverage distance is severely limited by the severe path loss. Thanks to its small wavelength, it can leverage large-scale antennas at transceivers to provide considerable beamforming gain to combat the path loss. Besides,  large-scale antennas at transceivers can increase spectral  efficiency (SE) via spatial multiplexing (SMX) \cite{Yu2016}.  For an mmWave multiple-input multiple-output (MIMO) system with $N_t$ transmit antennas, $N_r$ receive antennas and  $N_s$ data streams to be transmitted via $N_{RF}^t$ radio frequency (RF) chains, SMX over the best $N_s$-dimension ($N_s$-dim) beamspace, referred as the best beamspace selection (BBS) solution, has been regarded as the optimal solution and widely adopted as benchmarks in literature \cite{Ayach2014,Chen2015,Chen2017,Chen2018,Dai2015,Han2015,He2016,Gao2016,Li2017,Liang2014,Molu2018,Rusu2016,Yu2016,Park2017}.  The equivalent $N_s\times N_r$ MIMO  water-filling (WF-MIMO) channel capacity was treated as an unsurpassed SE upper bound.

Recently, researchers have proposed various SMX transmission schemes to approach the BBS solution and the corresponding performance bound \cite{Ayach2014,Chen2015,Chen2017,Chen2018,Dai2015,Han2015,He2016,Gao2016,Li2017,Liang2014,Molu2018,Rusu2016,Yu2016,Park2017,Liao2017,Mao2018}. In detail, BBS can be easily realized by fully-digital precoders. However, fully-digital precoders require costly RF chains with many signal mixers and analog-to-digital converters equipped comparable in number to the antenna elements \cite{Yu2016}. The cost and power consumption make fully-digital precoding infeasible in mmWave MIMO with large-scale antennas. To address the issue, hybrid digital and analog precoders  have been proposed \cite{Ayach2014,Chen2015,Chen2017,Chen2018,Dai2015,Han2015,He2016,Gao2016,Li2017,Liang2014,Molu2018,Rusu2016,Yu2016,Park2017}. For instance, to minimize the difference between the hybrid precoders and the fully-digital precoder, a spatially sparse precoding
scheme was proposed leveraging the sparsity of mmWave channel in \cite{Ayach2014}. It is based on a fully-connected structure. In  \cite{Yu2016}, more advanced alternating minimization algorithms were developed for the hybrid precoder designs, which achieve almost the same performance as the fully-digital procoder.  
By decomposing the hybrid precoding problem
into a series of sub-rate optimization problems, the authors in \cite{Dai2015} and \cite{Han2015} proposed  novel precoding algorithms based on successive interference cancellation (SIC). Recently, a hybrid precoding design with dynamic partially-connected structure has been investigated by trading off  complexity and SE  \cite{Park2017}.  Although the
hybrid precoding with either fully-connected or partially-connected structure can offer lower complexity and cost, it should be noticed that these
merits are achieved at the expense of  SE compared to BBS. In other words, they are all inferior to BBS in SE. As a result, they all treat BBS as an unsurpassed benchmark. But, is the SE of BBS truly supreme? Inspired by the concept of spatial
modulation (SM) and its variants \cite{Mesleh2008,Yang2008,Renzo2014,Ishikawa2018}, we propose a transmission scheme namely generalized beamspace modulation using multiplexing (GBMM) in this paper to challenge BBS in SE and the $N_s\times N_r$ WF-MIMO capacity.

\subsection{Prior Work}
 SM techniques have been recently introduced to mmWave MIMO systems. Prior work can be classified into two categories based on antenna switch and beamspace switch, respectively.

\subsubsection{SM techniques for mmWave MIMO based on antenna switch} The purest form of SM, namely space shift keying (SSK), was first applied to mmWave communications in \cite{Liu2015}. It showed that SSK can be applied to line-of-sight (LOS) mmWave MIMO communications as long as antennas are appropriately placed.  In \cite{Yang2017}, a receive antenna selection aided SM-MIMO scheme is proposed to reduce the number of the receive RF chains.   Quadrature spatial modulation (QSM) is a new SM MIMO technique to enhance the performance of SM while retaining the concept of antenna switching \cite{Mesleh2015}.  Capacity analysis
 for QSM based outdoor mmWave communications was presented in \cite{Younis2017}. Moreover, \cite{Ishikawa2017,Liu2016,Liu2018,He2017,He2018} have investigated generalized spatial modulation (GSM) at mmWave frequencies. The concept of GSM using multiplexing (GSMM) for hybrid precoding mmWave MIMO has been recently proposed in \cite{He2018}. 
All these works demonstrated that SM  techniques based on antenna switch could work effectively in mmWave communications. However, because of the antenna selection, most of the antennas are silent at each transmission slot and this reduces the beamforming gain compared with that fully utilizes all antennas.

\subsubsection{SM techniques for mmWave MIMO based on beamspace switch}
 Unlike above works, SM techniques based beamspace switch carry spatial domain information via the selection of beamforming vectors \cite{Perovic2017,Lee2017,Wang2018,Ding2017}. For instance, \cite{Perovic2017} proposed a receive spatial modulation (RSM) for LOS mmWave MIMO communications. However,  RSM was based on receive antenna selection and could not benefit from the array gain of the receive antennas. Focusing on improving error performance, the authors of \cite{Lee2017} proposed a virtual space modulation (VSM) transmission scheme and hybrid precoder designs  in mmWave MIMO. A maximum ratio combining (MRC)-based VSM transmission scheme was proposed and analog precoder design was investigated to enhance symbol error rate (SER) performance in \cite{Wang2018}. \cite{Ding2017} proposed a spatial scattering modulation (SSM) for uplink mmWave communications and studied its bit error rate (BER) performance. However, it should be noticed that all these works were done from the error performance perspective and with a single data stream.

In addition, very few literature on SM techniques based mmWave MIMO compares their schemes with the benchmark BBS  in SE. In \cite{He2018}, the authors compared the proposed GSMM with $N_s\times N_r$ WF-MIMO capacity. However, they failed to outperform it. To the best of our knowledge, there is no literature reporting the superiority of their schemes based on SM over BBS or $N_s\times N_r$ WF-MIMO capacity.  Therefore, it is the first time to  challenge the well known ``best'' BBS solution in terms of SE in mmWave MIMO communications adopting SM.
\subsection{Contributions}
\begin{itemize}
\item This paper proposes a GBMM technique for mmWave MIMO. The GBMM scheme not only fully utilizes all antennas for beamforming gain but also utilizes multiple-dim beamspace to convey multiple data streams via SMX, as well as  the beamspace hopping to carry information, which results into an increase of SE. The increase of SE  is achieved without changing the transceiver structure. GBMM outperforms BBS that only relies on SMX over the best beamspace for data transmission. We prove that the SE of the proposed GBMM outperforms the $N_s\times N_r$ WF-MIMO capacity that existing schemes try to approach.

\item In most of the literature on SM techniques, equal-probability antenna/beamspace activation is investigated. Differently, beamspaces in the proposed GBMM are activated with a general probability distribution. There exists an optimal probability distribution \cite{Liu2016}. This paper analyzes the upper and lower bounds of the achievable SE with the designed precoders and the general activation probability distribution. The upper bound is tight in the high signal-to-noise-ratio (SNR) regime. The lower bound adding a constant can provide a close approximation of SE in the low and high SNR regime.

\item To maximize the SE lower bound, we formulate an optimization problem  finding the optimal precoders, the best precoder activation probability distribution as well as the optimal power allocation among parallel data streams. To solve the formulated problem, we propose a gradient ascent algorithm applicable in all SNR regimes, but with high computational complexity.  To gain more insights and reduce the complexity, we also investigate a problem to maximize the tight upper bound of SE in the high SNR regime. The optimal precoder design, precoder activation probability distribution and  power allocation are derived in closed-form expressions, which significantly reduce system implementation complexity.

\item We discuss the implementation issues of GBMM. Specifically, we formulate optimization problems for hybrid precoder design and hybrid combiner design. We propose a coding approach to realize the optimized precoder activation probability distribution. We discuss an extension to orthogonal frequency division multiplexing (OFDM)-based broadband mmWave MIMO. Results indicate that the proposed GBMM scheme can be directly extended to broadband systems to improve SE.
\end{itemize}

Compared to the conference submission \cite{Guo2019} which introduces GBMM to the community, this paper includes more insightful closed-form solutions of optimal activation probability, precoder design and power allocation. The receiver-aware transceiver design and the coding approach to realize the precoder activation and the extension to broad-band mmWave MIMO systems are also discussed in this paper.  
\subsection{Organization}
The rest of this paper is organized as follows. In Section II, we describe the system model and introduce the conventional BBS transmission solution as well as its SE upper bound. In this section, we also describe the proposed GBMM and prove its superiority. In Section III, we use mutual information to characterize the SE of the proposed GBMM. We formulate the problems finding optimal fully-digital precoders,  precoder activation probability, power allocation and hybrid precoders to maximize  SE.  In Section IV, we propose a gradient ascent algorithm to solve one of the formulated problems, within which the complexity analysis is included. In this section, we derive the optimal solution to maximize SE in the high SNR regime. In Section V,  practical limits, implementation issues and extension are discussed. Specifically,  the impact of receiver structure is probed. A coding approach to realize the optimized precoder activation probability distribution is introduced. Also, an extension to broadband systems is  discussed. In Section VI, simulation results are illustrated to validate all analysis. We conclude in the last section.

\subsection{Notation} We use the following notations throughout the paper. $\textbf{A}$ denotes a matrix. $\textbf{a}$ represents a vector. $a$ stands for a scaler. $\mathcal{A}$ is a set. $\textbf{A}^H$ denotes the conjugate transpose of $\textbf{A}$. $\det(\textbf{A})$, $\mathrm{rank}(\textbf{A})$ and $\tr{(\textbf{A})}$ represent the determinate, the rank and trace of matrix $\textbf{A}$, receptively. $|\mathcal{A}|$ represents the size of set $\mathcal{A}$. $||\textbf{A}||_{F}$ denotes the Frobenius norm of matrix $\textbf{A}$. $||\textbf{a}||_{0}$ denotes the $l_0$ norm of vector $\textbf{a}$. $\diag{(\textbf{A})}$ represents a vector formed by the diagonal elements of matrix $\textbf{A}$. $\diag{(\textbf{a})}$ stands for a diagonal matrix with diagonal elements being $\textbf{a}$.
We use $\log$ and $\ln$ to represent the  logarithm functions of base $2$ and $e$, respectively. $\mathcal{I}(\cdot,\cdot)$ represents the mutual information function. $\mathcal{H}(\cdot)$ stands for the entropy function. $\mathcal{CN}(\boldsymbol{\mu},\boldsymbol{\Sigma})$ represents a complex Gaussian vector with mean $\boldsymbol{\mu}$ and covariance $\boldsymbol{\Sigma}$. $\mathbb{C}$ represents the complex field. $\mathbb{R}$ denotes the real number field. $\otimes$ represents the Kronecker product between two matrix or two vectors. $\textbf{I}_N$ is the $N\times N$ identity matrix. $\textbf{0}_{m}$ and $\textbf{1}_n$ stand of all zeros vector of size $m$ and all ones vector of size $n$, repectively. Expectation is denoted by $\mathbb{E}(\cdot)$. $(x)^+$ represents the function $\max\{x,0\}$.  $\mathcal{U}_{M\times N} $ denotes the set of all $M\times N$ matrices that have unit magnitude entries.

\section{System Model}
\begin{figure}[t]
  \centering
  \includegraphics[width=0.55\textwidth]{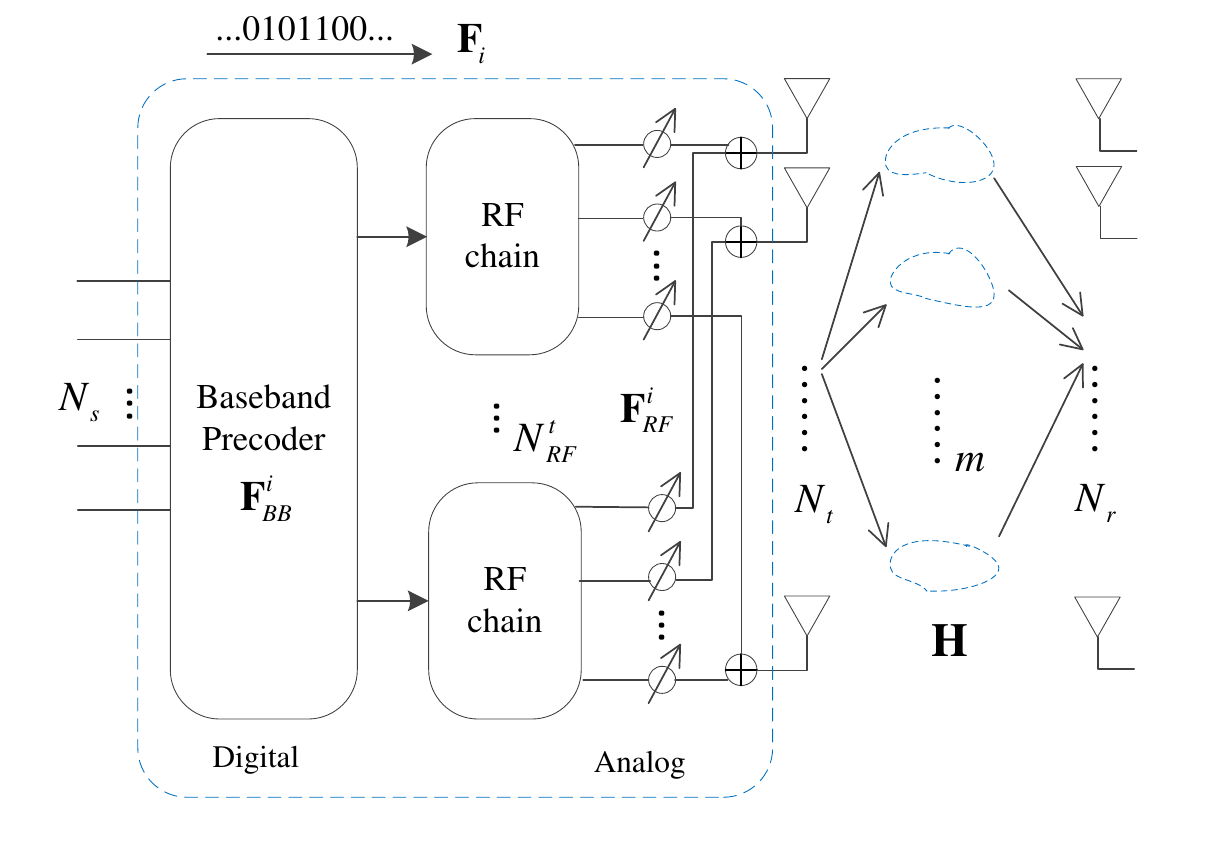}\\
 \caption{An ($N_t,N_r,N_{RF}^t,m$) mmWave MIMO system.}
  \label{Transmitter}
\end{figure}
We consider an $(N_t,N_r,N_{RF}^t,m)$ mmWave MIMO system, equipped with $N_{RF}^t(\ll N_t)$ RF chains as illustrated in Fig. \ref{Transmitter}, where $N_t$ and $N_r$ represent the number of transmit antennas and receive antennas, respectively. 
$\textbf{H}\in\mathbb{C}^{N_r\times N_t}$ is used to denote the channel matrix, which is perfectly known by the transceivers. $m$ denotes the rank of $\textbf{H}$ indicating the number of available parallel channels. Let $\textbf{x}\in\mathbb{C}^{N_t}$ represent the transmitted signal vector and the receive signal vector $\textbf{y}\in\mathbb{C}^{N_r}$ can be expressed as
\begin{equation}
\textbf{y}=\sqrt{\rho}\textbf{H}\textbf{x}+\textbf{n},
\end{equation}
where $\sqrt{\rho}$ represents the average receive SNR and  $\textbf{n}\in\mathbb{C}^{N_r}$ represents the noise vector with zero mean and unit variance, i.e., $\textbf{n}\sim\mathcal{CN}(\textbf{0}_{N_r},\textbf{I}_{N_r}\textbf)$.
 $\textbf{s}\in \mathbb{C}^{N_s}$ is used to represent the data symbol vector transmitted relying on the RF chains, where $N_s$ represents the number of data streams.
 It is assumed that the number of data streams and the number of RF chains  are less than the number of available parallel channels, i.e., $ N_s\leq N_{RF}^t <m$. The Saleh-Valenzuela channel model \cite{Yu2016} is adopted in this paper, which is expressed as
\begin{equation}
\begin{split}
\textbf{H}=\sqrt{\frac{N_tN_r}{N_{\mathrm{cl}}N_{\mathrm{ray}}}}\sum_{i=1}^{N_{\mathrm{cl}}}\sum_{l=1}^{N_{\mathrm{ray}}} \alpha_{il}
\textbf{b}_r(\phi_{il}^r,\theta_{il}^r) \textbf{b}_t(\phi_{il}^t,\theta_{il}^t)^H,
\end{split}
\end{equation}
where  $N_{\mathrm{cl}}$ represents the number of scattering clusters, $N_{\mathrm{ray}}$ denotes the number of propagation paths, $\alpha_{il}$ is the channel chain of $l$th ray in the $i$th propagation cluster,  $\phi_{il}^r$ and $\theta_{il}^r$ stand for azimuth and elevation angles of arrival (AoAs), respectively. $\phi_{il}^t$ and $\theta_{il}^t$ represents azimuth and elevation angles of departure (AoDs), respectively. It is assumed that $\{\alpha_{il}\}$ are independent and identically distributed (i.i.d.) random variables that follow the complex Gaussain distribution $\mathcal{CN}(0,\sigma_{\alpha,i}^2)$ and $\sum_{i=1}^{N_{\mathrm{cl}}}\sigma_{\alpha,i}^2=\beta$ which is the normalization factor to satisfy $\mathbb{E}\left[||\textbf{H}||_{F}^2\right]=N_tN_r$. In this paper, an $\sqrt{N}\times\sqrt{N}$ uniform square planar array (USPA) is considered, whose array response vector corresponding to the $l$th ray in the $i$th cluster writes
\begin{equation}\label{SteeringVector}
\begin{split}
\textbf{b}(\phi_{il},\theta_{il})&=\frac{1}{\sqrt{N}}\left[1,\cdots,e^{j\frac{2\pi}{\lambda}d(n_1\sin\phi_{il}\sin\theta_{il}+n_2\cos\theta_{il})},\right.
\\&\left.\cdots,e^{j\frac{2\pi}{\lambda}d((\sqrt{N}-1)\sin\phi_{il}\sin\theta_{il}+(\sqrt{N}-1)\cos\theta_{il})}\right]^T,
\end{split}
\end{equation}
where $\lambda$ and $d$ represent  the signal wavelength and the antenna spacing, respectively. In (\ref{SteeringVector}), $0\leq n_1< \sqrt{N}$ and $0\leq n_2<\sqrt{N}$ stand for the antenna indices in the two-dimensional (2D) plane. 

\subsection{Conventional Transmission Solution and Upper Bound}
Conventionally, a fixed precoder $\textbf{F}$ of dimension $N_t\times N_s$ is used to fit the transmission over $\textbf{H}$ with $||\textbf{F}||_F^2=N_s$. With $\textbf{F}$, the transmit vector $\textbf{x}$ can be expressed as $\textbf{x}=\textbf{Fs}$. The achievable SE can be characterized by the mutual information as
\begin{equation}\label{limit}
\begin{split}
\mathcal{R}(\textbf{H},\textbf{F})&=\mathcal{I}(\textbf{x};\textbf{y})=\mathcal{I}(\textbf{s};\textbf{y})\\
&=\log_2\det\left(\textbf{I}_{N_r}+\frac{\rho}{N_s}\textbf{H}\textbf{F}\textbf{F}^H\textbf{H}^H\right).
\end{split}
\end{equation}
To maximize the SE, $\textbf{F}$ is designed by solving 
\begin{equation}
\max_{||\textbf{F}||_F^2=N_s}\log_2\det\left(\textbf{I}_{N_r}+\frac{\rho}{N_s}\textbf{H}\textbf{F}\textbf{F}^H\textbf{H}^H\right).
\end{equation}
The BBS solution\footnote{The solution is obtained without taking the power allocation among data streams into consideration. Taking that into consideration, the solution should be $\textbf{F}^{\mathrm{opt}}=\textbf{V}_1\textbf{D}_{\mathrm{WF}}$, where $\textbf{D}_{\mathrm{WF}}\in\mathbb{C}^{N_r\times N_t}$ is the diagonal water-filling power allocation matrix with $\tr(\textbf{D}_{\mathrm{WF}}\textbf{D}_{\mathrm{WF}}^H)=N_s$.  The reason why the solution without power allocation is often considered in literature is that  $\textbf{D}_{\mathrm{WF}}^H$ is SNR dependent, while $\textbf{V}_1$ is SNR independent and applicable in all SNR regimes.  Moreover, $\textbf{D}_{\mathrm{WF}}$ approaches an identity matrix as SNR increases.} $\textbf{F}^{\mathrm{opt}}=\textbf{V}_1\in{\mathbb{C}^{N_t\times N_s}}$ is widely adopted, where $\textbf{V}_1$ is a matrix composed of $N_s$ right singular vectors of $\textbf{H}$ that correspond to the largest $N_s$  singular values. 
As is well known, the mutual information $\mathcal{I}(\textbf{s};\textbf{y})\leq\mathcal{C}^{\mathrm{WF}}_{N_s\times N_r}$, where $\mathcal{C}^{\mathrm{WF}}_{N_s\times N_r}$ represents the equivalent $N_s\times N_r$ WF-MIMO channel capacity. $\mathcal{C}^{\mathrm{WF}}_{N_s\times N_r}$ can be achieved by using BBS and water-filling power allocation. As aforementioned, this bound is well recognized as an unsurpassed benchmark of literature. But, is it true for  $(N_t,N_r,N_{RF}^{t},m)$ mmWave MIMO systems?
\subsection{The Proposed Transmission Solution and Its Superiority}
The real mutual information limit in (\ref{limit}) should be 
\begin{equation}
\begin{split}
\max_{f_{\textbf{x}}(\textbf{x})\atop\mathbb{E}\left(\mathcal{||\textbf{x}||}_{F}^2\right)=N_s}\mathcal{I}(\textbf{x};\textbf{y})&=\max_{f_{\textbf{s}|\textbf{F}}(\textbf{s}|\textbf{F}),~f_{\textbf{F}}(\textbf{F})\atop\mathbb{E}\left( \mathcal{||\textbf{Fs}||}_{F}^2\right)=N_s}\mathcal{I}(\textbf{F},\textbf{s};\textbf{y})\\& =\max_{f_{\textbf{s}|\textbf{F}}(\textbf{s}|\textbf{F}),~f_{\textbf{F}}(\textbf{F})\atop\mathbb{E}\left( \mathcal{||\textbf{Fs}||}_{F}^2\right)=N_s}\mathcal{I}(\textbf{s};\textbf{y})+\mathcal{I}(\textbf{F};\textbf{y}|\textbf{s})\\
& \geq \max_{f_{\textbf{s}}(\textbf{s})\atop\mathbb{E}\left(\mathcal{||\textbf{s}||}_{F}^2\right)=1}\mathcal{I}(\textbf{s};\textbf{y})=\mathcal{C}^{\mathrm{WF}}_{N_s\times N_r}.
\end{split}
\end{equation}
The exact solution to the problem is still unknown. But, it implies that we can additionally transmit information by making use of $\textbf{F}$ to improve the SE. Inspired by the concept of SM, we propose to use a general discrete distribution of $\mathcal{F}$ to enhance the SE as
\begin{equation}\label{FPDF}
f_{\textbf{F}}(\textbf{F})=P(\textbf{F}=\textbf{F}_i)=p_i,~ \textbf{F}_i\in\mathcal{F},
\end{equation}
where $\mathcal{F}$ denotes the set of candidate precoders.
(\ref{FPDF}) means that a precoder $\textbf{F}_i\in\mathcal{F}$ will be activated with probability $p_i$. For all $\textbf{F}_i\in\mathcal{F}$, we have $\sum_{i=1}^{|\mathcal{F}|}p_i=1$. As to the transmit data symbol vector $\textbf{s}$, we assume that it follows an i.i.d. complex Gaussian distribution independent of $\textbf{F}$ with zero mean and normalized power, i.e., $\textbf{s}\sim\mathcal{CN}(\textbf{0}_{N_s}, \frac{1}{N_s}\textbf{I}_{N_s})$. 
Defining $\textbf{p}\triangleq[p_1,p_2,\cdots,p_{|\mathcal{F}|}]^T$ and using $\mathcal{R}(\textbf{p},\mathcal{F})$ to represent the SE, we have the following theorem:
\begin{theorem}
\begin{equation}
\max_{\textbf{1}_{|\mathcal{F}|}^T\textbf{p}=1,~\textbf{p}\succeq \textbf{0}_{|\mathcal{F}|}\atop\mathbb{E}\left(||\textbf{F}||_{F}^2\right)=N_s,\textbf{F}\in\mathcal{F}}\mathcal{R}(\textbf{p},\mathcal{F})\geq \mathcal{C}^{\mathrm{WF}}_{N_s\times N_r}
\end{equation}
\end{theorem}
\begin{IEEEproof}
To prove, a set $\mathcal{F}$ that ensures the inequality is first proposed. Specifically,  $\textbf{F}_i\in \mathcal{F}$ is designed as
\begin{equation}\label{eqF}
\textbf{F}_i=\textbf{V}\textbf{E}_i\textbf{D}_i,
\end{equation}
where $\textbf{V}\in\mathbb{C}^{N_t\times m}$ represents the matrix composed of the right singular vectors of $\textbf{H}$  that correspond to all sorted  non-zero singular values;
$\textbf{E}_i\in\mathbb{C}^{m\times N_s}$ is a selection matrix to choose $N_s$ right-singular vectors  from all feasible $m$ ones. It is composed of a combination of $N_s$ base vectors. For instance,
\begin{equation}
\textbf{E}_1=[\textbf{e}_1,\textbf{e}_2,\cdots,\textbf{e}_{N_s}],
\end{equation}
where $\textbf{e}_i\in \mathbb{C}^{m}$ is the $i$th $m$-dim base vector with all zeros expect the $i$th entry being $1$. $\textbf{D}_i$ stands for the power allocation matrix and we have $\sum_{i=1}^{\mathcal{|F|}}p_i\tr(\textbf{D}_i\textbf{D}_i^H)=N_s$. In the design, $|\mathcal{F}|=\left(m\atop N_{s}\right)$. It is obvious that the BBS $\textbf{F}^{\mathrm{opt}}=\textbf{V}_1$ is  a special case of $\textbf{V}\textbf{E}_i$ because $\textbf{V}_1=\textbf{V}\textbf{E}_1$. By setting $\textbf{D}_1=\textbf{D}_{\mathrm{WF}}$ and $P(\textbf{F}=\textbf{F}_1)=1$, one can achieve $\mathcal{C}^{WF}_{N_s\times N_r}$. Thus, based on the fact that $\mathcal{C}^{WF}_{N_s\times N_r}$ is achieved by a specific realization of $\mathcal{F}$ and $\textbf{p}$, it is concluded that systems with globally optimized  $\mathcal{F}$ and $\textbf{p}$ will  achieve a larger SE than or the equal SE to $\mathcal{C}^{\mathrm{WF}}_{N_s\times N_r}$.
\end{IEEEproof}

\emph{Remark:} Based on above analysis in this section, we remark that  one can improve the SE of ($N_t$, $N_r$, $N_{RF}^t$, $m$) mmWave MIMO systems by additionally employing precoder (i.e., beamspace) selection  to carry information. As the beamspace indices are adopted as an additional modulation domain and multiple data streams are transmitted via SMX, we name the transmission solution as GBMM. In addition, $||\textbf{x}||_0=||\textbf{F}_i\textbf{s}||_0=N_t$ in the proposed GBMM scheme, indicating that all antennas are activated in a transmit slot. Compared with GSMM  \cite{He2018} based on antenna selection, GBMM will benefit from the beamforming gain by using all antennas. The approach is  attractive because the increase of SE is achieved without changing the transceiver structure. Besides, most reported SM techniques  and their various variants for all MIMO systems can be treated as specific realizations of $\mathcal{F}$ with $p_1=p_2=\cdots=p_{|\mathcal{F}|}=\frac{1}{|\mathcal{F}|}$ and $||\textbf{F}_1||_{F}^{2}=||\textbf{F}_2||_{F}^{2}=\cdots=||\textbf{F}_{|\mathcal{F}|}||_{F}^{2}=N_s$.

\section{SE Analysis and  Problem Formalization}
\subsection{Achievable SE}
We define two sets $\mathcal{E}\triangleq\{\textbf{E}_1,\textbf{E}_2,\cdots, \textbf{E}_{|\mathcal{F}|}\}$ and $\mathcal{D}\triangleq\{\textbf{D}_1,\textbf{D}_2,\cdots, \textbf{D}_{|\mathcal{F}|}\}$. Based on the design of $\textbf{F}_i\in\mathcal{F}$ in (\ref{eqF}), the variable set $\mathcal{F}$ in $\mathcal{R}(\textbf{p},\mathcal{F})$ can be replaced by $\mathcal{D}$, because $\mathcal{E}$ is known. 
According to the similar analysis in \cite{Ibrahim2016}, the achievable SE $\mathcal{R}(\textbf{p},\mathcal{D})$ can be derived as
\begin{equation}\label{ExactSE}
\begin{split}
\mathcal{R}(\textbf{p},\mathcal{D})&=\mathcal{H}(\textbf{y})-\mathcal{H}(\textbf{y}|\textbf{H},\textbf{x})\\
&=\mathbb{E}[-\log f(\textbf{y})]-N_r\log(\pi e)\\
&=-\int_{\mathbb{C}^{N_r}}\log f(\textbf{y})\sum_{i=1}^{|\mathcal{F}|}p_i f_i(\textbf{y}) dy-N_r\log(\pi e),
\end{split}
\end{equation}
where 
\begin{equation}
f(\textbf{y})=\sum_{i=1}^{|\mathcal{F}|}p_i f_i(\textbf{y}),
\end{equation}
\begin{equation}
f_i(\textbf{y})=\frac{1}{\pi^{N_r}\det(\boldsymbol{\Sigma}_i)}\exp\left(-\textbf{y}^H\boldsymbol{\Sigma}_i^{-1}\textbf{y}\right),
\end{equation}
\begin{equation}
\boldsymbol{\Sigma_i}=\frac{\rho}{N_s}\textbf{H}\textbf{V}\textbf{Q}_i\textbf{V}^H\textbf{H}^H+\textbf{I}_{N_r},
\end{equation}
and 
\begin{equation}
\textbf{Q}_i=\textbf{E}_i\textbf{D}_i\textbf{D}_i^H\textbf{E}_i^H.
\end{equation}
The exact SE has more theoretical significance than practical significance since it involves the integration of a complicated function thus not offering clear insights into the system performance.
 To gain insights and to provide practical system design guidelines,  the following upper bound and lower bound of SE are used:
\begin{proposition}
The achievable SE $\mathcal{R}(\textbf{p},\mathcal{D})$ is upper bounded by
\begin{equation}\label{eqUB}
\mathcal{R}^{U}(\textbf{p},\mathcal{D})=\sum_{i=1}^{|\mathcal{F}|}p_i\left(\log \det \boldsymbol{\Sigma}_i-\log p_i\right),
\end{equation}
and lower bounded by
\begin{equation}\label{eqLB}
\mathcal{R}^{L}(\textbf{p},\mathcal{D})=-\sum_{i=1}^{|\mathcal{F}|}p_i\log\left(\sum_{j=1}^{|\mathcal{F}| }p_j z_{i,j}\right)-N_r\log e,
\end{equation}
where $z_{i,j}=1/\left[\det(\boldsymbol{\Sigma}_i+\boldsymbol{\Sigma}_j)\right]$. 
The upper bound is tight in the high SNR regime and the lower bound adding a constant gap $N_r(\log e-1)$ is tight in the low and high SNR regime. 
\end{proposition}
\begin{IEEEproof}
The proofs of the bounds and the tightness can be found in \cite{Ibrahim2016} and \cite{He2018}, respectively.
To avoid repetition, we omit them for brevity. 
\end{IEEEproof}

\subsection{Problem Formulation}

We aim to maximize the achievable SE, which is a maximization problem. For the sake of low complexity, the lower bound in (\ref{eqLB}) can be adopted as the objective function.  The optimization problem is hence formulated as 
\begin{equation}
\begin{split}
(\textbf{P1}): \max~~ &\mathcal{R}^{L}(\textbf{p},\mathcal{D})\\
\mathrm{subject~to:}~~&
\textbf{1}_{|\mathcal{F}|}^T\textbf{p}=1,~\textbf{p}\succeq \textbf{0}_{|\mathcal{F}|},\\
&\sum_{i=1}^{\mathcal{|F|}}p_i\tr(\textbf{D}_i\textbf{D}_i^H)=N_s.
\end{split}
\end{equation}
The optimized $\textbf{p}$ and $\mathcal{D}$ yielded by solving ($\textbf{P1}$) are applicable in all SNR regimes. However, since the expression of SE lower bound in (\ref{eqLB}) is still too complicated, it is difficult to obtain a closed-form solution. To solve ($\textbf{P1}$), we have to resort to a numerical optimization approach, which will render too much complexity. To gain insights and to reduce the implementation complexity, the maximization of the upper bound in (\ref{eqUB}) is also investigated in this paper, because the upper bound is tight in the high SNR regime and given by a simpler expression. The corresponding optimization problem is formulated as
\begin{equation}
\begin{split}
(\textbf{P2}): \max~~ &\mathcal{R}^{U}(\textbf{p},\mathcal{D})\\
\mathrm{subject~to:}~~&
\textbf{1}_{|\mathcal{F}|}^T\textbf{p}=1,~\textbf{p}\succeq \textbf{0}_{|\mathcal{F}|},\\
&\sum_{i=1}^{\mathcal{|F|}}p_i\tr(\textbf{D}_i\textbf{D}_i^H)=N_s.
\end{split}
\end{equation}
Using the optimized $\mathcal{D}$, we can directly construct $\mathcal{F}$ based on (\ref{eqF}).  The costly fully-digital structure is required  to implement the $\textbf{F}_i$ in $\mathcal{F}$. To reduce the implementation cost and complexity,  we design hybrid digital and analog precoders by solving
\begin{equation}
\begin{split}
(\textbf{P3}):\min~~&||\textbf{F}_i-\textbf{F}_{RF}^{i}\textbf{F}_{BB}^{i}||_{F}^2\\
\mathrm{subject~to:}~~& \textbf{F}_{RF}^{i}\in \mathcal{U}_{N_t\times N_{RF}^t},~\textbf{F}_{BB}^i\in \mathbb{C}^{N_{RF}^{t}\times N_s},\\
&|| \textbf{F}_{RF}^{i}\textbf{F}_{BB}^{i}||_F^2=|| \textbf{F}_{i}||_F^2,
\end{split}
\end{equation}
where $\textbf{F}^{i}_{RF}$ and $\textbf{F}^{i}_{BB}$ represent the $i$th analog precoder and digital precoder respectively and  $i=1,2,\cdots,|\mathcal{F}|$.
($\textbf{P3}$)  has been well investigated in a large body of literature \cite{Ayach2014,Chen2015,Chen2017,Chen2018,Dai2015,Han2015,He2016,Gao2016,Li2017,Liang2014,Molu2018,Rusu2016,Yu2016,Park2017}. Existing low-complexity algorithms in \cite{Ayach2014,Chen2015,Chen2017,Chen2018,Dai2015,Han2015,He2016,Gao2016,Li2017,Liang2014,Molu2018,Rusu2016,Yu2016,Park2017} can be directly applied to  solve $(\textbf{P3})$. Thus in the following of this paper, we focus on solving  problems  ($\textbf{P1}$) and ($\textbf{P2}$).

\section{Transmitter Optimization}

In this section, we find the optimal precoder activation probability distribution $\textbf{p}$ and  power allocation matrix set $\mathcal{D}$ (equivalent to designing $\mathcal{F}$) to maximize the bounds of the achievable SE.

\subsection{Optimization Applicable in All SNR Regimes}

We define $\textbf{D}_i\triangleq \diag(d_{i1},d_{i2},\cdots,d_{iN_s})$
, $\lambda_{ij}\triangleq d_{ij}^2$, $\boldsymbol{\lambda_i}\triangleq [\lambda_{i1},\lambda_{i2},\cdots,\lambda_{iN_s}]^T\in \mathbb{R}^{N_s}$,  $\boldsymbol{\lambda}\triangleq\left[\boldsymbol{\lambda_1}^T,\boldsymbol{\lambda_2}^T,\cdots, \boldsymbol{\lambda_2}^T\right]^T\in \mathbb{R}^{|\mathcal{F}|N_s}$ and $\textbf{q}\triangleq\textbf{p}\otimes \textbf{1}_{N_s}\in \mathbb{R}^{|\mathcal{F}|N_s}$. Based on these definitions, the problem ($\textbf{P1}$) becomes 
\begin{equation}
\begin{split}
(\textbf{P1a}): \max~~ &\mathcal{R}^{L}(\textbf{p},\boldsymbol{\lambda})\\
\mathrm{subject~to:}~~&
\textbf{1}_{|\mathcal{F}|}^T\textbf{p}=1,~\textbf{p}\succeq \textbf{0}_{|\mathcal{F}|},\\
&\textbf{q}^T\boldsymbol{\lambda}=N_s,~\boldsymbol{\lambda}\succeq \textbf{0}_{|\mathcal{F}|N_s}.
\end{split}
\end{equation}
To release the non-negative constraints $\textbf{p}\succeq \textbf{0}_{|\mathcal{F}|}$ and $\boldsymbol{\lambda}\succeq \textbf{0}_{|\mathcal{F}|N_s}$ in (\textbf{P1a}), we adopt a barrier method with the barrier function  defined as \cite{He2018}
\begin{equation}
\psi(u)=\left\{
\begin{array}{rcl}
\frac{1}{t_B}\ln u,      &     u>0\\
 -\infty,   &       u\leq 0,
\end{array} \right. 
\end{equation}
and rewrite the optimization problem as
\begin{equation}
\begin{split}
(\textbf{P1b}): \max~~ &f(\textbf{p},\boldsymbol{\lambda})\\
\mathrm{subject~to:}~~&
\textbf{1}_{|\mathcal{F}|}^T\textbf{p}=1,~\textbf{q}^T\boldsymbol{\lambda}=N_s,
\end{split}
\end{equation}
where $t_B$ is a factor to scale the barrier function's penalty\footnote{With larger $t_B$, $f(\textbf{p},\boldsymbol{\lambda})$ is closer to $\mathcal{R}^{L}(\textbf{p},\boldsymbol{\lambda})$.} and  the  objection function $f(\textbf{p},\boldsymbol{\lambda})$ is given by
\begin{equation}
f(\textbf{p},\boldsymbol{\lambda})=\mathcal{R}^{L}(\textbf{p},\boldsymbol{\lambda})+\sum_{i=1}^{|\mathcal{F}|}\psi(p_i)+\sum_{i=1}^{|\mathcal{F}|}\sum_{j=1}^{N_s}\psi(\lambda_{ij}).
\end{equation}
The gradient of $f(\textbf{p},\boldsymbol{\lambda})$ with respect to $\textbf{p}$ can be derived as
\begin{equation}\label{eqGP}
\begin{split}
\nabla_{\textbf{p}}&f(\textbf{p},\boldsymbol{\lambda})=[\nabla_{p_1}f(\textbf{p},\boldsymbol{\lambda}),\nabla_{p_2}f(\textbf{p},\boldsymbol{\lambda}),\cdots,\nabla_{p_{|\mathcal{F}|}}f(\textbf{p},\boldsymbol{\lambda})]^T,
\end{split}
\end{equation}
where 
\begin{equation}\label{eq_pi}
\nabla_{p_i}f(\textbf{p},\boldsymbol{\lambda})=\nabla_{p_i}\mathcal{R}^{L}(\textbf{p},\boldsymbol{\lambda})+\frac{1}{t_B}p_i^{-1},
\end{equation}
and
\begin{equation}
\begin{split}
\nabla_{p_i}{R}^{L}&(\textbf{p},\boldsymbol{\lambda})=
\\&-\frac{\rho\log e}{N_s}\left[\log\left(\sum_{j=1}^{|\mathcal{F}| }p_j z_{i,j}\right)+\sum_{k=1}^{|\mathcal{F}|}\frac{p_k z_{k,i}}{\sum_{j=1}^{|\mathcal{F}| }p_j z_{k,j}}\right].
\end{split}
\end{equation}
The gradient of $f(\textbf{p},\boldsymbol{\lambda})$ with respect to $\boldsymbol{\lambda}$ can be expressed as
\begin{equation}\label{eqGLambda}
\begin{split}
\nabla_{\boldsymbol{\lambda}}f(\textbf{p},\boldsymbol{\lambda})=[\nabla_{\boldsymbol{\lambda}_1}f(\textbf{p},\boldsymbol{\lambda}),\nabla_{\boldsymbol{\lambda}_2}f(\textbf{p},\boldsymbol{\lambda}),\cdots,\nabla_{\boldsymbol{\lambda}_{|\mathcal{F}|}}f(\textbf{p},\boldsymbol{\lambda})]^T.
\end{split}
\end{equation}
where 
\begin{equation}\label{lambda_i}
\nabla_{\boldsymbol{\lambda}_i}f(\textbf{p},\boldsymbol{\lambda})=\nabla_{\boldsymbol{\lambda}_i}\mathcal{R}^{L}(\textbf{p},\boldsymbol{\lambda})+\frac{1}{t_B}\boldsymbol{\delta}_i,
\end{equation}
\begin{equation}
\boldsymbol{\delta}_i\triangleq\left[\lambda_{i1}^{-1},\lambda_{i2}^{-1},\cdots,\lambda_{iN_s}^{-1}\right]^T,
\end{equation}
\begin{equation}
\begin{split}
\nabla_{\boldsymbol{\lambda}_i}{R}^{L}&(\textbf{p},\boldsymbol{\lambda})=
\\&-\log e\left[\sum_{k=1}^{|\mathcal{F}|}\frac{p_ip_k\nabla_{\boldsymbol{\lambda}_i} z_{i,k}}{\sum_{j=1}^{|\mathcal{F}|}p_jz_{i,j}}+\sum_{k=1,\atop k\neq i}^{|\mathcal{F}|}\frac{p_ip_k\nabla_{\boldsymbol{\lambda}_i} z_{k,i}}{\sum_{j=1}^{|\mathcal{F}|}p_jz_{k,j}}\right],
\end{split}
\end{equation}
and
\begin{equation}
\begin{split}
\nabla_{\boldsymbol{\lambda}_i} z_{i,k}
=\frac{\rho z_{i,k}}{N_s}\diag[\textbf{E}_i^H\textbf{V}^H\textbf{H}^H(\boldsymbol{\Sigma}_i+\boldsymbol{\Sigma}_k)^{-1}\textbf{H}\textbf{V}\textbf{E}_i].
\end{split}
\end{equation}

To meet the equality constraints $\textbf{1}_{|\mathcal{F}|}^T\textbf{p}=1$ and $\textbf{q}^T\boldsymbol{\lambda}=N_s$,
we perform the following projections
\begin{equation}\label{eqPP}
\Delta\textbf{p}=\left(\textbf{I}_{|\mathcal{F}|}-\frac{\textbf{1}_{|\mathcal{F}|}\cdot \textbf{1}_{|\mathcal{F}|}^T}{|\mathcal{F}|}\right)\nabla_{\textbf{p}}f(\textbf{p},\boldsymbol{\lambda}),
\end{equation}
\begin{equation}\label{eqPLambda}
\Delta\boldsymbol{\lambda}=\left(\textbf{I}_{|\mathcal{F}|N_s}-\frac{\textbf{1}_{|\mathcal{F}|N_s}\cdot \textbf{q}^T}{N_s}\right)\nabla_{\boldsymbol{\lambda}}f(\textbf{p},\boldsymbol{\lambda}),
\end{equation}
to ensure
\begin{equation}\label{c1}
\textbf{1}_{|\mathcal{F}|}^T\Delta\textbf{p}=0,
\end{equation}
and 
\begin{equation}\label{c2}
\textbf{q}^T\Delta\boldsymbol{\lambda}=0.
\end{equation}
Based on these gradients and projections, we develop a gradient ascent algorithm to maximize the SE lower bound as listed in \textbf{Algorithm 1}.

\begin{algorithm}[t]
\caption{Gradient ascent algorithm to maximize the SE lower bound}
\label{alg:A}
\begin{algorithmic}[1]
\STATE {Initialize $\textbf{p}^{(0)}$ with $p_i=\frac{1}{|\mathcal{F}|},~i=1,2,\cdots,|\mathcal{F}|$, $\boldsymbol{\lambda}^{(0)}$ with $\lambda_{ij}=1,~j=1,2,\cdots,N_s$, the halting criterion $\epsilon$ and the iteration number $k=0$.}
\REPEAT
\STATE{ Compute the gradient $\nabla_{\textbf{p}}f(\textbf{p},\boldsymbol{\lambda})$ by using (\ref{eqGP}). }
\STATE{Carry out the projection to get $\Delta\textbf{p}^{(k)}$ by using (\ref{eqPP}).}
\STATE{Solve the following problem via backtracking line search \cite{Boyd2004} and update}
\begin{equation*}
\eta_1^*=\arg \max_{\eta_1} f(\textbf{p}^{(k)}+\eta_1\Delta\textbf{p}^{(k)},\boldsymbol{\lambda}^{(k)}),
\end{equation*}
\begin{equation*}
\textbf{p}^{(k+1)}\leftarrow\textbf{p}^{(k)}+\eta_1^*\Delta\textbf{p}^{(k)},
\end{equation*}
\begin{equation*}
\textbf{q}^{(k+1)}=\textbf{p}^{(k+1)}\otimes \textbf{1}_{N_s}.
\end{equation*}
\STATE{Modify $\boldsymbol{\lambda}^{(k+1)}$ to ensure $\textbf{q}^T\boldsymbol{\lambda}=N_s$ as}
\begin{equation*}
\boldsymbol{\lambda}^{(k+1)}\leftarrow\frac{\boldsymbol{\lambda}^{(k)}}{\left(\textbf{q}^{(k+1)}\right)^T\boldsymbol{\lambda}^{(k)}}.
\end{equation*}
\STATE{Compute the gradient $\nabla_{\boldsymbol{\lambda}}f(\textbf{p},\boldsymbol{\lambda})$ by using (\ref{eqGLambda}). }
\STATE{Carry out the projection to get $\Delta\boldsymbol{\lambda}^{(k+1)}$ by using (\ref{eqPLambda}).}
\STATE{Solve the following problem via backtracking line search \cite{Boyd2004} and update}
\begin{equation*}
\eta_2^*=\arg \max_{\eta_2} f(\textbf{p}^{(k+1)},\boldsymbol{\lambda}^{(k+1)}+\eta_2\Delta\boldsymbol{\lambda}^{(k+1)}),
\end{equation*}
\begin{equation*}
\boldsymbol{\lambda}^{(k+1)}\leftarrow\boldsymbol{\lambda}^{(k+1)}+\eta_2^*\Delta\boldsymbol{\lambda}^{(k+1)},
\end{equation*}
\begin{equation*}
k\leftarrow k+1.
\end{equation*}
\UNTIL{$\eta_1^*||\Delta\textbf{p}^{(k)}||_{2}\leq\epsilon||\textbf{p}^{(k)}||_2$ and $\eta_2^*||\Delta\boldsymbol{\lambda}^{(k)}||_{2}\leq\epsilon||\boldsymbol{\lambda}^{(k)}||_2$}
\STATE{Output the optimized $\textbf{p}^*$ and use  $\boldsymbol{\lambda}^*$ to obtain $\mathcal{D}^*$ and $\mathcal{F}^*$.} 
\end{algorithmic}
\end{algorithm}
%

\emph{Complexity and Converge Analysis:} The complexity of \textbf{Algorithm 1} is dominated by the calculation of  the gradients in (\ref{eqGP}) and (\ref{eqGLambda}). In each iteration, the calculation of the gradient with respect to $\textbf{p}$ needs $|\mathcal{F}|^3$ calculations of matrix determinant of size $N_r\times N_r$. The calculation of the gradient with respect to $\boldsymbol{\lambda}$ requires $|\mathcal{F}|^3$ calculations of matrix determinant, matrix inversion, and matrix multiplication of size $N_r\times N_r$. Thus, the overall complexity is around $\mathcal{O}(|\mathcal{F}|^3N_r^3)$. The complexity is a heavy burden for large-scale antennas.  The algorithm will converge because the value of the objective function will increase in each iteration and it is upper bounded.

\emph{Gradient Modification to Avoid Convergence to Local Optima:}  
If there are more than one entries approaching zeros in the optimal $\textbf{p}^*$ or $\boldsymbol{\lambda}^*$, the algorithm may converge to a local optimum because the searching steps $\eta_1$ and $\eta_2$ in \textbf{Algorithm 1} will be  small if some entries of $\textbf{p}$ and $\boldsymbol{\lambda}$ approach zero in the searching procedure. To avoid this, we introduce a small threshold $\tau$. If any entries are smaller than $\tau$,  the projected gradients to them are forced to be zeros. To ensure (\ref{c1}) and (\ref{c2}), the projection of gradients should be modified. Specifically, taking the optimization of $\textbf{p}$ as an example, the projected gradient vector is modified as
\begin{equation}
\Delta \textbf{p}_{\mathcal{K}_{-}}=\textbf{0}_{|\mathcal{K}_{-}|},
\end{equation}
and 
\begin{equation}\label{eqPPp}
\Delta\textbf{p}_{\mathcal{K}_{+}}=\left(\textbf{I}_{|\mathcal{K}_{+}|}-\frac{\textbf{1}_{|\mathcal{K}_{+}|}\cdot \textbf{1}_{|\mathcal{K}_{+}|}^T}{|\mathcal{K}_{+}|}\right)\nabla_{\textbf{p}_{\mathcal{K}_{+}}}f(\textbf{p},\boldsymbol{\lambda}).
\end{equation}
where $\mathcal{K}_+$ and $\mathcal{K}_-$ denote the index sets that $\textbf{p}_{\mathcal{K}_+}\succeq  \tau\textbf{1}_{|\mathcal{K}_+|}$ and $\textbf{p}_{\mathcal{K}_-}\prec  \tau\textbf{1}_{|\mathcal{K}_-|}$, respectively.
In this way, not only can \textbf{Algorithm 1} converge to a global optimum, but also greatly reduce the computational complexity, because we do not need to compute $\nabla_{\textbf{p}_{\mathcal{K}_{-}}}f(\textbf{p},\boldsymbol{\lambda})$. Similarly, such a technique can be adopted in the optimization of $\boldsymbol{\lambda}$.
 The performance and convergence of \textbf{Algorithm 1} with/without the gradient modification are investigated in Section VI.
\subsection{Optimization Only Applicable in the High SNR Regime}

\begin{theorem}
For a fixed $\mathcal{D}$,  the optimal activation probability distribution in the high SNR regime is
\begin{equation}\label{eqpi}
p_i^{\star}=\frac{2^{c_i}}{\sum_{i=1}^{|\mathcal{F}|}2^{c_i}}, ~i=1,2,\cdots, |\mathcal{F}|,
\end{equation}
where $c_i\triangleq\log \det  \boldsymbol{\Sigma}_i$ is the capacity of the $N_s\times N_r$ MIMO channels when $\textbf{F}_i$ is activated.
\end{theorem}

\begin{IEEEproof}
Lagrange function of the problem $\textbf{P2}$ can be formulated as
\begin{equation}\label{eqL1}
\begin{split}
J(\textbf{p},\mu)&=\mathcal{R}^{U}(\textbf{p})-\mu(\sum_{i=1}^{|\mathcal{F}|}p_i-1)\\
&=\sum_{i=1}^{|\mathcal{F}|}p_i\left( c_i-\log p_i\right)-\mu(\sum_{i=1}^{|\mathcal{F}|}p_i-1).
\end{split}
\end{equation}
 Taking the  partial derivation of the Lagrange function in (\ref{eqL1}) with respect to $p_i$, we deduce a set of equations as
\begin{equation}
 c_i-\log p_i-1/\ln2 -\mu=0,~i=1,2,\cdots,|\mathcal{F}|,
\end{equation}
 based on which and the constraint $\sum_{i=1}^{|\mathcal{F}|}p_i-1=0$, the optimal $p_i^{\star}$ in (\ref{eqpi}) is derived. 
\end{IEEEproof}
\begin{lemma}
Using a fixed $\textbf{p}$ for transmission, the optimal power allocated the $j$th data stream when  $\textbf{F}_i$ is activated in the high SNR regime  is obtained as
\begin{equation}\label{power_allocation}
\lambda_{i,j}=\left(\frac{1}{\xi_i\ln2}-\frac{1}{\rho\sigma_{ij}^2}\right)^+,
\end{equation}
and $\xi_i$ is obtained by solving
\begin{equation}
\sum_{j=1}^{N_s}\left(\frac{ 1}{\xi_{i}\ln2}-\frac{1}{\rho\sigma_{ij}^2}\right)^+=b_i,
\end{equation}
and
\begin{equation}
\sum_{i=1}^{|\mathcal{F}|}p_ib_i=N_s,
\end{equation}
where $b_i$ is the average power when $\textbf{F}_i$ is activated.
\end{lemma}
\begin{IEEEproof}
To prove, we also use the same definitions of $\lambda_{ij}$ and $\boldsymbol{\lambda}$ as in Section IV-A. Additionally, we use $\boldsymbol{\sigma}_i\triangleq[\sigma_{i1},\sigma_{i2},\cdots,\sigma_{iN_s}]^T$ to denote the $N_s$ singular values selected from all $m$ ones of $\textbf{H}$ by using the selection matrix $\textbf{E}_i$.
Based on the definitions, $\mathcal{R}^{U}(\boldsymbol{\lambda})$ can be expressed as 
\begin{equation}
\mathcal{R}^{U}(\boldsymbol{\lambda})=\sum_{i=1}^{|\mathcal{F}|}\sum_{j=1}^{N_s}p_i\log\frac{\rho\sigma_{ij}^2\lambda_{ij}+1}{p_i}.
\end{equation}
Lagrange function can be formulated to be
\begin{equation}\label{eqL2}
\begin{split}
\mathcal{R}^{U}(\boldsymbol{\lambda},\xi)=&\sum_{i=1}^{|\mathcal{F}|}\sum_{j=1}^{N_s}p_i\log\frac{\rho\sigma_{ij}^2\lambda_{ij}+1}{p_i}\\&-\sum_{i=1}^{|\mathcal{F}|}\xi_ip_i\left(\sum_{j=1}^{N_s}\lambda_{ij}-b_i\right).
\end{split}
\end{equation}
Taking the partial derivation of the Lagrange function in (\ref{eqL2}) with respect to $\lambda_{ij}$, we obtain a set of equations as
\begin{equation}
\frac{\rho\sigma_{ij}^2p_i}{\ln2(\rho\sigma_{ij}^2\lambda_{ij}+1)}-\xi_i p_i=0,~i=1,\cdots,|\mathcal{F}|,~j=1,\cdots, N_s.
\end{equation}
Solving the equations and $\sum_{i=1}^{|\mathcal{F}|}\sum_{j=1}^{N_s}p_i\lambda_{ij}=N_s$, we can obtain the optimized power allocation in \textbf{Lemma 1}.
\end{IEEEproof}

\emph{Remark:} The optimal precoder activation probability distribution $\textbf{p}$ and power allocation matrix $\mathcal{D}$ (i.e. equivalent to designing $\mathcal{F}$) are derived by fixing either one in the high SNR regime. It can be used to verify whether the numerical approach \textbf{Algorithm 1} is able to obtain a best solution in the high SNR regime. We note that the optimal power allocation in $\textbf{Lemma 1}$ is the famous water-filling power allocation when $\{b_i\}$, $i=1,2,\cdots, |\mathcal{F}|$ are fixed. Next, we need to find the best $\textbf{b}\triangleq\left[b_1,b_2,\cdots, b_{|\mathcal{F}|}\right]$.

\begin{lemma}
In the high SNR regime, the optimal solution of $\textbf{b}$ is $b_1=b_2=\cdots=b_{|\mathcal{F}|}=N_s$.
\end{lemma}
\begin{IEEEproof}
For the equivalent MIMO channels when $\textbf{F}_i$ is activated, its capacity in the high SNR regime can be expressed as \cite{Goldsmith2003}
 \begin{equation}
c_i=N_s\log b_i+g_i,
\end{equation} 
 where $g_i$ is a constant related to the channel gains. Therefore, the optimization problem of finding $\textbf{b}$ so as to maximize $\mathcal{R}^U(\textbf{b})$ can be written as
\begin{equation}\label{Pbi}
\begin{split}
\max~&\mathcal{R}^U(\textbf{b})=\sum_{i=1}^{|\mathcal{F}|}p_i\left[ N_s\log b_i+g_i-\log p_i\right]\\
&\mathrm{subject~to}~\sum_{i=1}^{|\mathcal{F}|}p_ib_i=N_s.
\end{split}
\end{equation}
Lagrange function of the problem in (\ref{Pbi}) is formulated as
\begin{equation}\label{eqL3}
\begin{split}
J(\textbf{b},\zeta)
=\sum_{i=1}^{|\mathcal{F}|}p_i\left[ N_s\log b_i+g_i+\log p_i\right]-\zeta(\sum_{i=1}^{|\mathcal{F}|}p_ib_i-N_s).
\end{split}
\end{equation}
 Taking the  partial  derivation of the Lagrange function in (\ref{eqL3}) with respect to $b_i$,  a set of equations are obtained as
\begin{equation}
\frac{N_sp_i}{b_i}-\zeta p_i=0, i=1,\cdots,|\mathcal{F}|,
\end{equation}
based on which and the constraint $\sum_{i=1}^{|\mathcal{F}|}p_ib_i-N_s=0$,  the optimal solution as $b_1=b_2=\cdots=b_{|\mathcal{F}|}=N_s$ is obtained. 
\end{IEEEproof}
Based on \textbf{Lemma 1} and \textbf{Lemma 2}, one can easily derive the following theorem:

\begin{theorem}
The optimal power allocation is independent of the activation probability distribution $\textbf{p}$ in the high SNR regime, which can be expressed as
\begin{equation}
\lambda_{i,j}=\left(\frac{1}{\xi_i^\star\ln2}-\frac{1}{\rho\sigma_{ij}^2}\right)^+,
\end{equation}
where $\xi^\star_i$  satisfies
\begin{equation}
\sum_{j=1}^{N_s}\left(\frac{ 1}{\xi_{i}^{\star}\ln2}-\frac{1}{\rho\sigma_{ij}^2}\right)^+=N_s.
\end{equation}
\end{theorem}
\begin{IEEEproof}
The optimal power allocation is straightforward to be derived by substituting $b_1=b_2=\cdots=b_{|\mathcal{F}|}=N_s$ (\textbf{Lemma 2}) into (\ref{power_allocation}) (\textbf{Lemma 1}). 
\end{IEEEproof}

Based on \textbf{Theorem 2} and \textbf{Theorem 3}, we can obtain the optimal solution to $(\textbf{P2})$ as listed in \textbf{Algorithm 2}.

\begin{algorithm}[t]
\caption{Optimal solution to maximize the SE upper bound}
\label{alg:A}
\begin{algorithmic}[1]
\STATE{Compute the optimal power allocation $\boldsymbol{\lambda}^{\star}$ by using water-filling algorithm according to \textbf{Theorem 3} as
\begin{equation*}
\lambda_{i,j}^{\star}=\left(\frac{1}{\xi_i^\star\ln2}-\frac{1}{\rho\sigma_{ij}^2}\right)^+,
\end{equation*}
where $\xi^\star_i$ satisfies}
\begin{equation}
\sum_{j=1}^{N_s}\left(\frac{ 1}{\xi_{i}^{\star}\ln2}-\frac{1}{\rho\sigma_{ij}^2}\right)^+=N_s.
\end{equation}
\STATE {Use  $\boldsymbol{\lambda}^{\star}$ to generate $\mathcal{D}^{\star}$, $\mathcal{F}^{\star}$ and compute  $\textbf{p}^{\star}$ according to \textbf{Theorem 2}:\\
~~~Compute $c_i,~i=1,2,\cdots, |\mathcal{F}|$ based on $\boldsymbol{\lambda}$ and}
\begin{equation*}
p_i^{\star}= \frac{2^{c_i}}{\sum_{i=1}^{|\mathcal{F}|}2^{c_i}}, ~i=1,2,\cdots, |\mathcal{F}|.
\end{equation*}

\STATE{Output the optimized $\textbf{p}^{\star}$, $\boldsymbol{\lambda}^{\star}$, $\mathcal{D}^{\star}$ and $\mathcal{F}^{\star}$.} 
\end{algorithmic}
\end{algorithm}
\emph{Complexity Analysis:} The complexity of $\textbf{Algorithm 2}$ is dominated by the complexity of the water-filling power allocation for $|\mathcal{F}|N_s$ parallel channels and the computation of  $c_i,~i=1,2,\cdots, |\mathcal{F}|$. Among them, the complexity of  power allocation for $|\mathcal{F}|N_s$ parallel channels  is $\mathcal{O}(|\mathcal{F}|N_s)$ by using the linear-complexity water-filling algorithm \cite{Khakurel2014}. The computation of  $c_i^{(k)},~i=1,2,\cdots, |\mathcal{F}|$ requires $|\mathcal{F}|$ calculations of matrix determinate of size $N_r\times N_r$. The complexity is  $\mathcal{O}(|\mathcal{F}|N_r^3)$.  
As $N_s\ll N_r$, it is concluded that the overall complexity of \textbf{Algorithm 2} is $\mathcal{O}(|\mathcal{F}|N_r^3)$.  Moreover, no iteration is needed. For clearly viewing the complexity of \textbf{Algorithm 1} and \textbf{Algorithm 2} in finding the optimized $\textbf{p}$, $\mathcal{D}$ and $\mathcal{F}$ in $(N_t,N_r,N_{{RF}}^t,m)$ mmWave MIMO systems, we list the analytical results in Table I where $N_{iter}$ is the number of iterations that \textbf{Algorithm 1} takes to converge and $|\mathcal{F}|=\left(m\atop N_s\right)$.
\begin{table}[t]
\centering
\caption{The Complexity of Algorithm 1 and Algorithm 2}\label{tab1}
    \begin{tabular}{ | c | c |}
    \hline
 Algorithm & Complexity\\
\hline
\textbf{Algorithm 1} & $\mathcal{O}\left(N_{iter}\left(m\atop N_s\right)^3N_r^3\right)$\\
\hline
\textbf{Algorithm 2} & $\mathcal{O}\left(\left(m\atop N_s\right)N_r^3\right)$\\
\hline
    \end{tabular}
\end{table}

\section{Discussion and Extension}
\subsection{Practical Limit and Implementation Challenges}
\begin{figure}[t]
  \centering
  \includegraphics[width=0.55\textwidth]{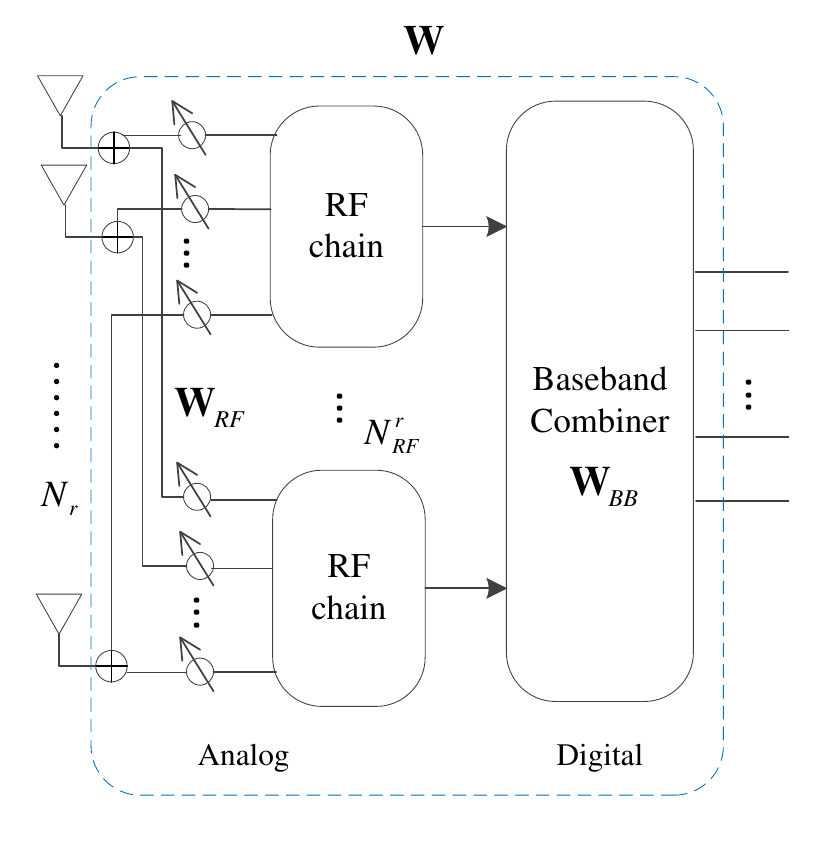}\\
 \caption{Hybrid digital and analog receiver structure}
  \label{Receiver}
\end{figure}
The achievable SE of the proposed GBMM scheme is greatly affected by the receiver structure. To investigate the impact, a receiver with $N_{RF}^r$  RF chains as depicted in Fig. \ref{Receiver} is considered in this paper. We remark as follows.
\begin{proposition}
To benefit from GBMM, the number of equipped receive  RF chains should be greater than the number of data streams $N_s$, i.e., $N_{RF}^r>N_s$ .
\end{proposition}
\begin{IEEEproof}
 We use $\textbf{W}\in \mathbb{C}^{N_r\times\hat{m}}$ to represent the receive combiner,  where $\hat{m}\leq N_{RF}^r$.  With the combiner, the equivalent channels can be expressed as $\tilde{\textbf{H}}=\textbf{W}^H\textbf{H}$. Let $\tilde{m}$  be the rank of $\tilde{\textbf{H}}$ and we have $\tilde{m}=\mathrm{rank}(\tilde{\textbf{H}})=\min{(\hat{m},m)}$. The precoders can be designed as $\tilde{\textbf{F}}_i=\tilde{\textbf{V}}\tilde{\textbf{E}}_i\tilde{\textbf{D}}_i$ with the precoder set size  $\tilde{|\mathcal{F}|}=\left(\tilde{m}\atop N_s\right)$, where $\tilde{\textbf{V}}\in\mathbb{C}^{N_t\times \tilde{m}}$ denotes the matrix composed of the right-singular vectors of $\tilde{\textbf{H}}$ corresponding to all non-zero singular values; $\tilde{\textbf{E}}_i\in\mathbb{C}^{\tilde{m}\times N_s}$, $\tilde{\textbf{D}}_i\in\mathbb{C}^{N_s\times N_s}$ represent beamspace selection matrix and power allocation matrix, respectively.  To benefit from the information carrying capability of precoder selection, $\tilde{m}=\min{(\hat{m},m)}$ should be larger than $N_s$ .  Thus, not only $m$ but also $\hat{m}$ must be greater than $N_s$. As $\hat{m}\leq N_{RF}^r$, we deduce $ N_{RF}^r>N_s$.  This is reasonable because as more information is conveyed by precoder selection over $N_s$ independent data streams, a higher degree is required.
\end{IEEEproof}

In this paper, we adopt $\textbf{W}=\tilde{\textbf{U}}$, where $\tilde{\textbf{U}}$ is an $N_r\times\hat{m}$ matrix composed of the $\hat{m}$ left-singular vectors of $\tilde{\textbf{H}}$ that correspond to the largest $\hat{m}$  singular values.
Based on the optimal receiver,  the digital combiner $\textbf{W}_{BB}\in \mathbb{C}^{  N_{RF}^{r}\times \tilde{m}}$ and analog combiner $\textbf{W}_{RF}\in \mathcal{U}_{ N_r\times N_{RF}^r }$ can be designed by
\begin{equation}
\begin{split}
(\textbf{P4}):\min_{\textbf{W}_{RF},\textbf{W}_{BB}}&~||\textbf{W}-\textbf{W}_{RF}\textbf{W}_{BB}||_{F}^2\\
\mathrm{subject~to:}&~\textbf{W}_{RF}\in \mathcal{U}_{ N_r\times N_{RF}^r},~\textbf{W}_{BB}\in \mathbb{C}^{  N_{RF}^{r}\times \tilde{m}}.\\
\end{split}
\end{equation}
The problem ($\textbf{P4}$) is similar to  ($\textbf{P3}$). Both of them can be solved by existing algorithms developed in  \cite{Ayach2014,Chen2015,Chen2017,Chen2018,Dai2015,Han2015,He2016,Gao2016,Li2017,Liang2014,Molu2018,Rusu2016,Yu2016,Park2017}.

To implement GBMM, it mainly faces two challenges. The first one is the heavy computational complexity in solving problems (\textbf{P1})-(\textbf{P4}). For instance, for the hybrid precoder design in $(\textbf{P3})$, we have to run existing algorithms in \cite{Ayach2014,Chen2015,Chen2017,Chen2018,Dai2015,Han2015,He2016,Gao2016,Li2017,Liang2014,Molu2018,Rusu2016,Yu2016,Park2017} for $|\mathcal{F}|$ times. Therefore,  more efficient lower-complexity algorithms dedicated to the hybrid precoder designs of GBMM are still in demand. It is possible because $\textbf{F}_i\in\mathcal{F}$ are related to each other. The other way to reduce the complexity is by reducing the size of the set $\mathcal{F}$. This can be done by forcing the activation probability under a threshold to be zero. For instance, if we want $|\mathcal{F}|\leq 10$, we can set activation probability below $0.1$ to be $0$. The second implementation challenge faced with is the switching speed of precoders. The switch operation has to be done in a symbol duration.  Therefore, the switching speed of precoders, especially the analog precoders using phase shifters, is a key factor that determines the data rate of GBMM. The authors of \cite{Wang2018} have researched the switching speed of analog phase shifters. Specifically, \cite{Romanofsky2007} compared four kinds of phase shifters including semiconductor, ferroelectric, ferrite,
and micro-electromechanical systems (MEMS),
which showed that the switching speed of semiconductor and ferroelectric is in order of  the nanosecond. \cite{Co2006} introduced a novel low-cost  phase shifters, whose switching speed is in the range of tens of nanoseconds. Thanks to these high switching-speed phase shifters, it is feasible to implement high-rate GBMM.

\subsection{Coding Approach to Realize the Optimized Precoder Activation Probability Distribution}
Most of SM techniques consider equal-probability antenna/precoder activation by using uniformly distributed information bits to activate the antenna  directly. It should be noted that even with such kind of approach, the index activation probability distribution can be designed as non-uniform as shown in literature \cite{Guo2016,Guo2017,Wang2016,Wang2017}.
In this paper, we propose a general coding approach to approximately realize the optimized precoder activation probability distribution \textbf{p}. In the approach, the information of amount $\mathcal{H}(\textbf{p})$ is coded with more bits. It is assumed that the number of coded bits is $n_b$. The $n_b$-bit codewords are divided into $|\mathcal{F}|$ groups with group index mapping to the precoder index. The ratio of group sizes over $2^{n_b}$ is set to be approximately equal to the distribution $\textbf{p}$. In this way,  the information of  amount $\mathcal{H}(\textbf{p})$ can be conveyed in a codeword. The coding rate can be expressed as $r_b=\log\mathcal{H}(\textbf{p})/n_b$.
\subsection{Extension to Broadband}
The paper is considered in the narrow band to gain insights regarding SE. With OFDM, the proposed GBMM scheme can be directly extended to broadband systems. In detail, the receive signal of the frequency domain can be expressed as
\begin{equation}
\textbf{y}[k]=\rho\textbf{H}[k]\textbf{F}_i[k]\textbf{s}[k]+\textbf{n}[k], k=1,2,\cdots,K,
\end{equation}
where $k$ represents the sub-carrier index and $K$ is the number of carriers.
The fully-digital precoder design and precoder activation probability distribution can be optimized  as same as that in the narrow band except that either ($\textbf{P1}$) or ($\textbf{P2}$) should be solved for $K$ times. As the transmissions over all sub-carriers rely on the same analog precoders, the hybrid precoder  design problems become
\begin{equation}
\begin{split}
(\textbf{P5}):\sum_{k=1}^{k}\min~~&||\textbf{F}_i[k]-\textbf{F}_{RF}^{i}\textbf{F}_{BB}^{i}[k]||_{F}^2\\
\mathrm{subject~to:}~~& \textbf{F}_{RF}^{i}\in \mathcal{U}_{N_t\times N_{RF}^t},~\textbf{F}_{BB}^i[k]\in \mathbb{C}^{N_{RF}^{t}\times N_s},\\
&|| \textbf{F}_{RF}^{i}\textbf{F}_{BB}^{i}[k]||_F^2=|| \textbf{F}_{i}[k]||_F^2.
\end{split}
\end{equation}
The algorithms for solving (\textbf{P5}) can also be found in rich literature such as \cite{Yu2016}, \cite{Ayach2014} and \cite{Sohrabi2016}.

\section{Simulation and Analysis}
In this section, we  show the effectiveness as well as the convergence  of $\textbf{Algorithm 1}$ and numerically evaluate the SE performance of the proposed GBMM. It is assumed that both transmitter and  receiver are equipped with USPA. Specifically, the transmitter is with $N_t=100$ antennas and the receiver is with $N_r=36$ antennas. The channels are set as $N_{\mathrm{cl}}=4$ clusters, $N_{\mathrm{ray}}=2$ and the average power of each clusters, $\sigma_{\alpha，i}^2=1$. The azimuth and elevation AoAs and AoDs follow the Laplacian distribution, which have uniformly distributed mean angles over $[0,2\pi)$ and $10$ degrees angular spread as that in \cite{Yu2016}. The minimum antenna spacing in the USPA is a half wavelength. Despite this specific channel model is used in simulations, it should be noted that the proposed designs are applicable to more general models.
\subsection{Effectiveness and Convergence of Algorithm 1}
\begin{figure}[t]
  \centering
  \includegraphics[width=0.55\textwidth]{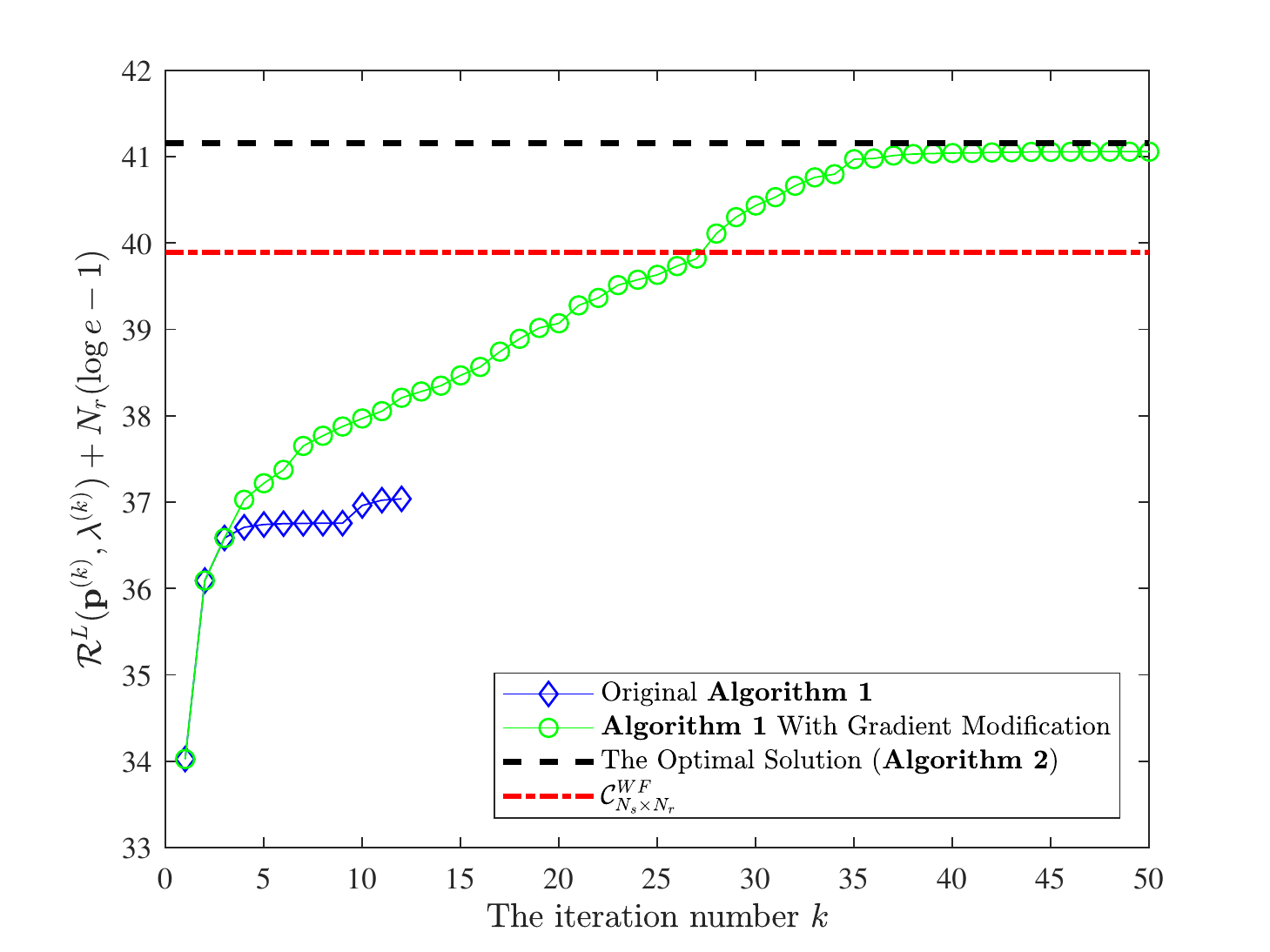}\\
 \caption{The performance and convergence of Algorithm 1 under a given channel realization at a high SNR of $15$ dB where $N_t=100$, $N_r=36$ and $N_{RF}^t=N_s=3$.}
  \label{result0}
\end{figure}
Firstly, to show the effectiveness as well as the convergence of \textbf{Algorithm 1}, we evaluate the performance of $\textbf{p}^{(k)}$ and $\boldsymbol{\lambda}^{(k)}$ of the iteration $k$ at a high SNR of $15$ dB. Instead of evaluating the object function $f(\textbf{p}^{(k)},\boldsymbol{\lambda}^{(k)})$, we calculate the values of more meaningful $\mathcal{R}^{L}(\textbf{p}^{(k)},\boldsymbol{\lambda}^{(k)})$ adding a constant $N_r(\log e-1)$, because these values provide a tight approximation of the exact SE. In the simulation, a channel realization is adopted and $N_{RF}^t=N_s=3$. 
The halting $\epsilon$ is set to $10^{-3}$ and the gradient modification threshold $\tau$ is $2\times10^{-3}$. For comparison, we also demonstrate the optimal solution given by $\textbf{Algorithm 2}$ and $\mathcal{C}_{N_s\times N_r}^{WF}$ (achieved by BBS with water-filling power allocation) under the same channel realization.
Results demonstrate that without modifying the gradients, the original $\textbf{Algorithm 1}$ converges fast but at a local optimum. The reason has been discussed in Section IV-A. With gradient modification described in Section IV-A, the issue can be well addressed, and the algorithm will finally converge to the global optimum given by $\textbf{Algorithm 2}$. Comparison with $\mathcal{C}_{N_s\times N_r}^{WF}$   shows that GBMM outperforms $\mathcal{C}_{N_s\times N_r}^{WF}$ more than $1$ bit/s/Hz under the  given channel realization.

\subsection{SE Evaluation}
\begin{figure}[t]
  \centering
  \includegraphics[width=0.55\textwidth]{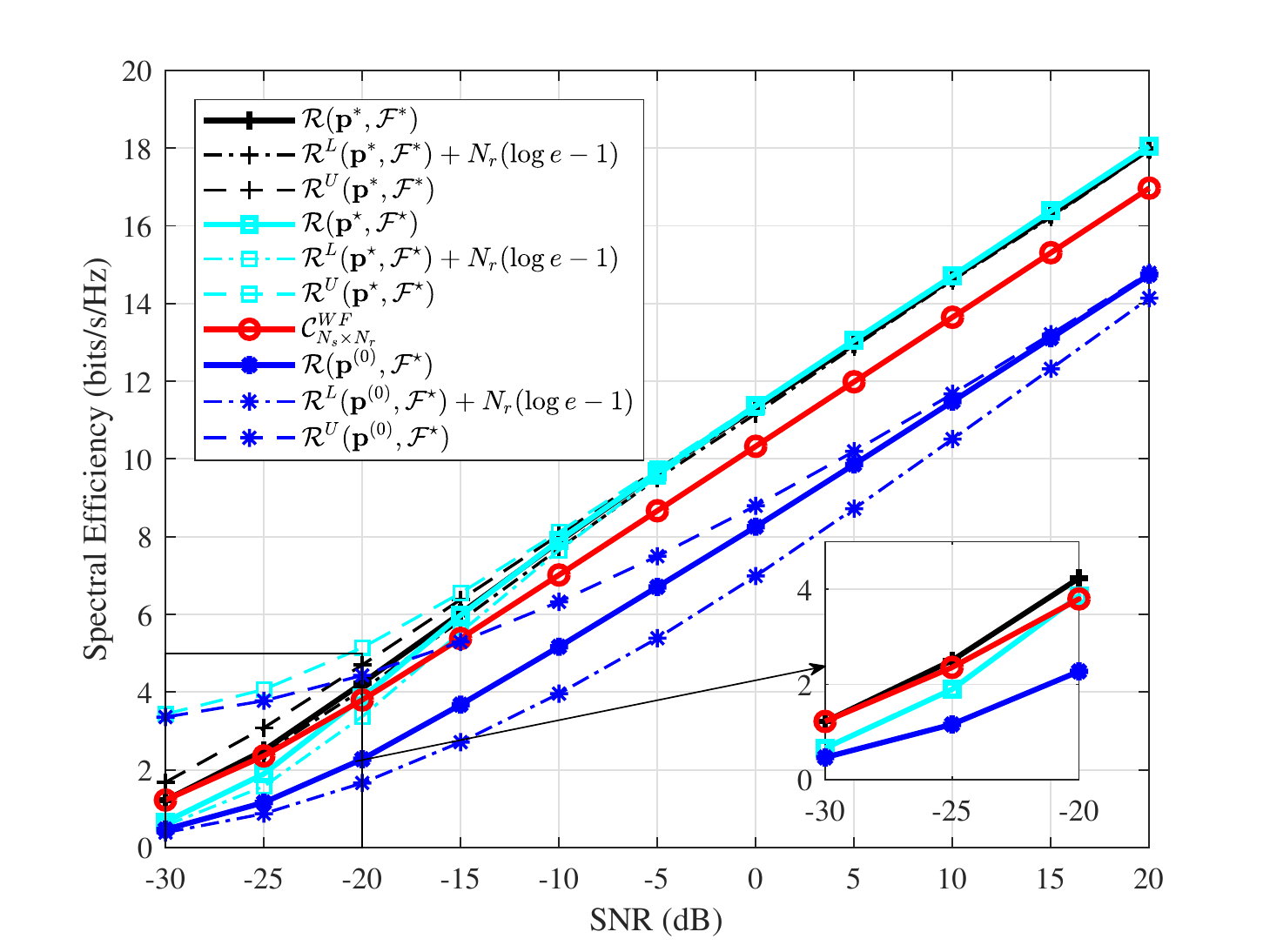}\\
 \caption{The SE of GBMM mmWave MIMO with different $\textbf{p}$, $\mathcal{F}$ where $N_t=100$, $N_r=36$ and $N_{RF}^t=N_s=1$.}
  \label{result1a}
\end{figure}

\begin{figure}[t]
  \centering
  \includegraphics[width=0.55\textwidth]{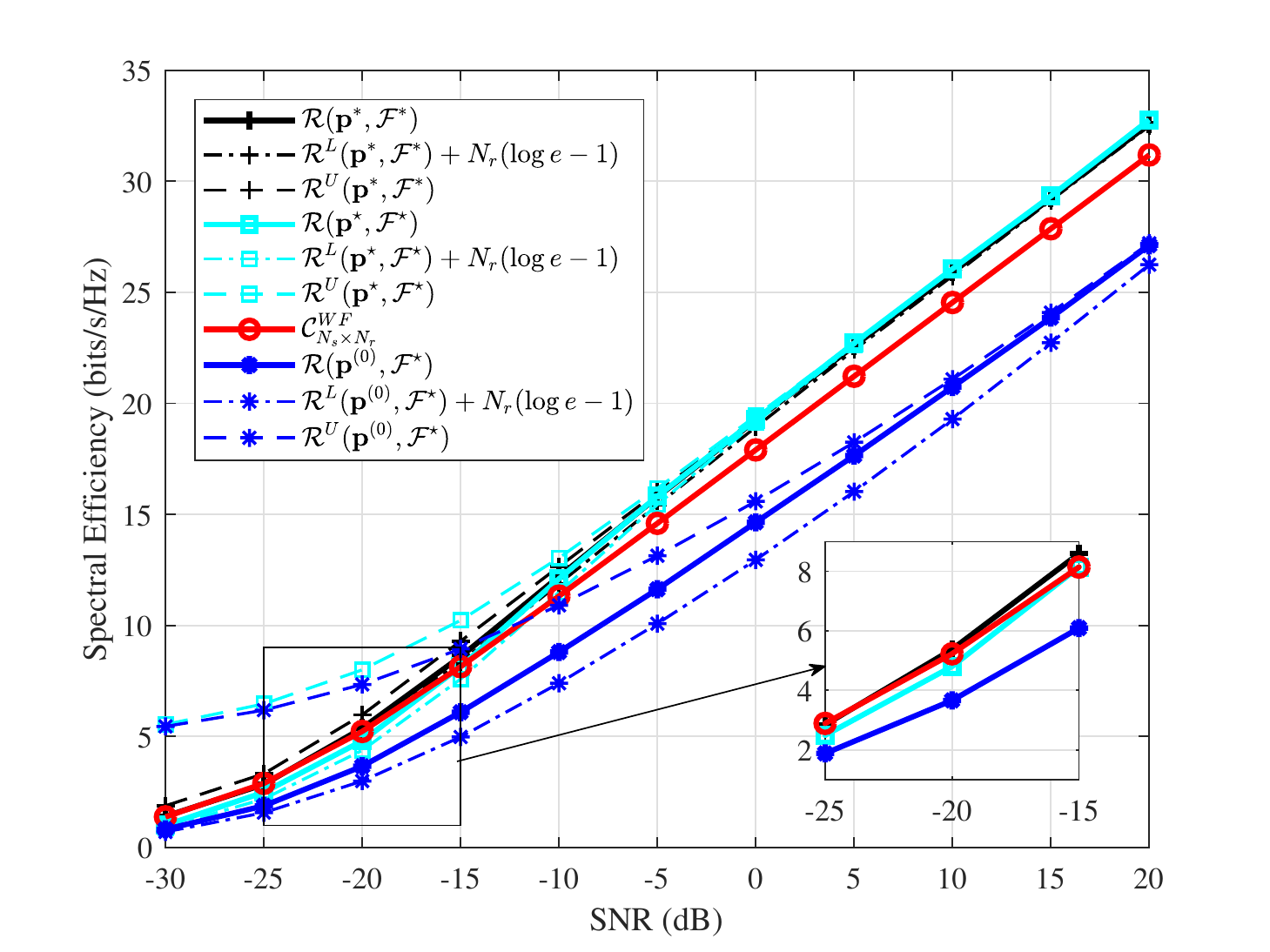}\\
 \caption{The SE of GBMM mmWave MIMO with different $\textbf{p}$, $\mathcal{F}$ where $N_t=100$, $N_r=36$ and $N_{RF}^t=N_s=2$.}
  \label{result1b}
\end{figure}
Secondly, we investigate the SE of the proposed GBMM scheme with different optimized fully-digital precoder sets $\mathcal{F}$ and precoder activation probability  distributions $\textbf{p}$ with $N_{RF}^t=N_s=1$ and $N_{RF}^t=N_s=2$ as illustrated in Fig. \ref{result1a} and Fig.~\ref{result1b}, respectively.  Specifically, three transmission solutions are considered. The first one is with the optimized  $\mathcal{F}^*$ and  $\textbf{p}^*$ obtained from \textbf{Algorithm 1} to maximize the SE lower bound. The second one is with the optimized $\mathcal{F}^{\star}$ and  $\textbf{p}^{\star}$ obtained from \textbf{Algorithm 2} to maximize the SE upper bound. The third one with the optimized $\mathcal{F}^{\star}$ and the equal-probability distribution $\textbf{p}^{(0)}$, which is widely adopted in  \cite{Perovic2017,Lee2017,Ding2017}.
We illustrate not only  the exact SE expressed in (\ref{ExactSE}) but also  the upper bound as well as the approximation (i.e., the lower bound adding a constant gap $N_r(\log e-1)$).
 For comparison, $\mathcal{C}_{N_s\times N_r}^{WF}$ is also included. All simulation results are averaged over $100$ channel realizations.

Results in Figs. 4-5 show the tightness of the upper bound in the high SNR regime and that of the lower bound  adding the constant $N_r(\log e-1)$ in the low and high SNR regime. It is demonstrated  that the numerical optimization method proposed in this paper (i.e., \textbf{Algorithm 1})  can approach the closed-form optimal solution (i.e., \textbf{Algorithm 2}) in the high SNR regime for all channel realizations. System with $(\textbf{p}^*,\mathcal{F}^*)$ obtained from \textbf{Algorithm 1} achieves the optimal SE performance and is superior to $\mathcal{C}_{N_s\times N_r}^{WF}$  over all depicted SNR regime and considerably outperforms $\mathcal{C}_{N_s\times N_r}^{WF}$ by more than $1$ bit/s/Hz in the high SNR regime. However, the complexity burden to get the solution is huge as discussed in Section IV-A. System with $(\textbf{p}^{\star},\mathcal{F}^{\star})$ adopting \textbf{Algorithm 2} achieves the highest SE performance in the high SNR regime and also considerably outperforms $\mathcal{C}_{N_s\times N_r}^{WF}$ by more than $1$ bit/s/Hz in the high SNR regime. It is slightly worse than the solution offered by \textbf{Algorithm 1} and $\mathcal{C}_{N_s\times N_r}^{WF}$  in the low SNR regime as observed in the enlarged figure. As this solution is obtained with much lower complexity than  \textbf{Algorithm 1}, it is more suitable for piratical implementations.
 Using the conventional SM concept with equal activation probability distribution $\textbf{p}^{(0)}$, the SE is much lower than the proposed GBMM solution with optimized $\textbf{p}^*$ or $\textbf{p}^\star$, which indicates the significance of precoder activation probability distribution in the optimization. Additionally, conventional SM with equal probability achieves the lowest SE  and can not outperform $\mathcal{C}_{N_s\times N_r}^{WF}$,  because its large activation probability of low-capacity channels will degrade the overall SE. However, in our design, the high-capacity channels are activated with high probability, and the low-capacity channels are activated with low probability as specified in $\textbf{Theorem 2}$. $\mathcal{C}_{N_s\times N_r}^{WF}$ achieves the highest SE in the low SNR regime, but it is no longer the highest in medium and high regime because BBS activates the channel with the highest capacity with probability $1$ and the amount of information carried on precoder index is, therefore, $0$.

\subsection{Hybrid Precoder Designs}
\begin{figure}[t]
  \centering
  \includegraphics[width=0.55\textwidth]{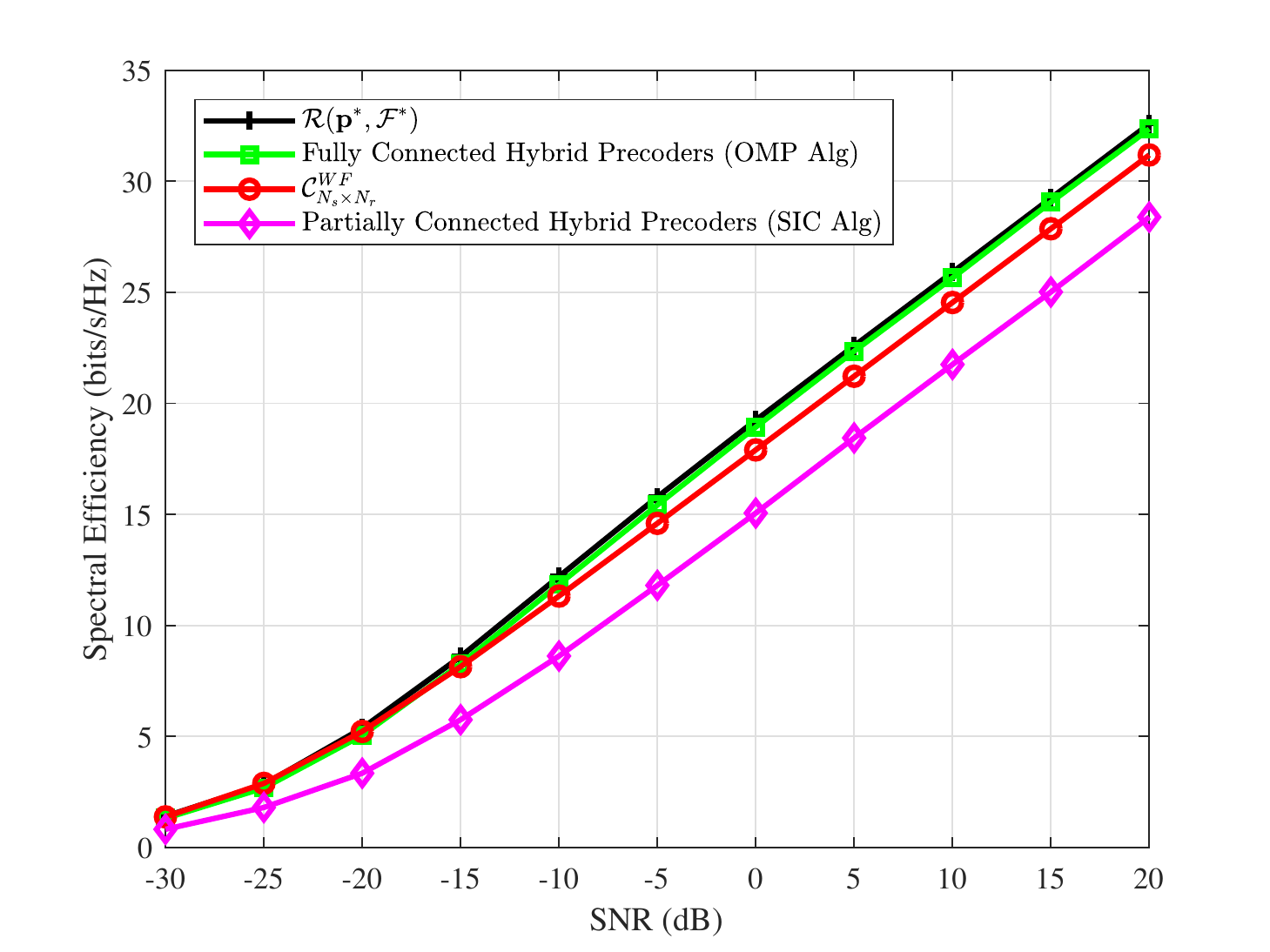}\\
 \caption{The SE of GBMM mmWave MIMO with hybrid precoders where $N_t=100$, $N_r=36$ and $N_{RF}^t=N_s=2$.}
  \label{result2}
\end{figure}

Thirdly, we investigate the hybrid precoder designs with either fully-connected structure or partially-connected structure as illustrated in Fig. \ref{result2}.  Existing algorithms are used to solve ($\textbf{P3}$). Specifically, we use OMP-Alg \cite{Ayach2014} for  fully-connected  hybrid precoder designs and SIC-Alg \cite{Dai2015} for partially-connected hybrid precoder designs.  In the simulations, $N_{RF}^{t}=N_s=2$ and all simulation results are averaged over $100$ channel realizations. For comparison,  $\mathcal{C}_{N_s\times N_r}^{WF}$ is also included as benchmark.
Results in Fig. \ref{result2} demonstrate that GBMM with fully-connected hybrid precoders has a small performance gap with the fully-digital precoders (i.e., $\mathcal{R}(\textbf{p}^{*},\mathcal{F}^{*})$) and outperforms $\mathcal{C}_{N_s\times N_r}^{WF}$. GBMM with partially-connected hybrid precoders will lead to a great SE loss. However, we note that the performance with hybrid precoders can be improved by utilizing more advanced precoder design algorithms \cite{Yu2016}. $\mathcal{R}(\textbf{p}^*,\mathcal{F}^*)$ is the new bound that all algorithms are unable to surpass. 

\begin{figure}[t]
  \centering
  \includegraphics[width=0.55\textwidth]{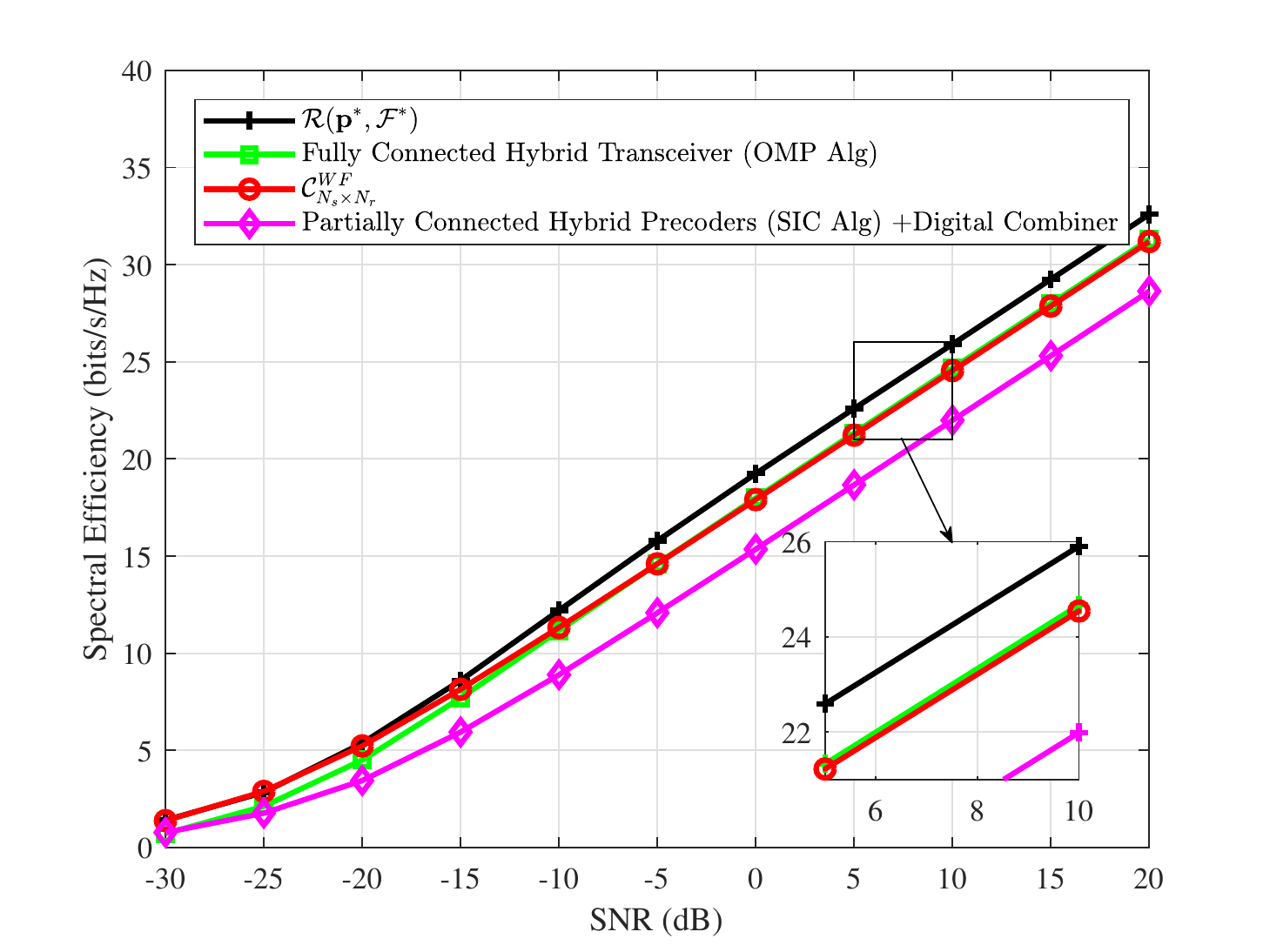}\\
 \caption{The SE of GBMM mmWave MIMO with hybrid transceivers where $N_t=100$, $N_r=36$, $N_{RF}^t=N_s=2$ and $N_{RF}^r=4$.}
  \label{result3}
\end{figure}

\subsection{Hybrid Transceiver Designs}
Thirdly, taking the impact of the hybrid receiver structure into consideration, we investigate the hybrid transceiver designs discussed in Section V. The simulation results are as illustrated in Fig. \ref{result4}. All simulation results are averaged over $100$ channel realizations. In the simulations, $N_{RF}^{r}=4$, $N_{RF}^t=N_s=2$.  We utilize the $\tilde{m}$ singular vectors that correspond the  largest $\tilde{m}$ singular values  to obtain $\tilde{\textbf{V}}$ and design  $\tilde{\mathcal{F}}^*$ and $\textbf{p}^*$ by using \textbf{Algorithm 1}.  We solve ($\textbf{P3}$) and ($\textbf{P4}$) by using  OMP-Alg in \cite{Ayach2014} for  fully-connected  hybrid transceiver designs. Moreover, we use the SIC-Alg in \cite{Dai2015} for partially-connected hybrid precoder designs and use a fully-digital combiner.
For comparison, the SE of fully-digital transceiver design (i.e., $\mathcal{R}(\textbf{p}^{*},\mathcal{F}^{*})$)  and $\mathcal{C}_{N_s\times N_r}^{WF}$ are also included. It is shown in Fig. \ref{result3} that the hybrid transceiver designs will degrade the SE as expected. Moreover, results also demonstrate the proposed GBMM with fully-connected hybrid transceiver can achieve slightly higher SE than the BBS solution with water-filling power allocation, i.e., $\mathcal{C}_{N_s\times N_r}^{WF}$, showing the potential of the proposed GBMM in improving SE in practical systems. Similarly, the performance of the hybrid transceiver designs can be improved with more advanced algorithms.

\subsection{GBMM for mmWave MIMO-OFDM Systems}
\begin{figure}[t]
  \centering
  \includegraphics[width=0.55\textwidth]{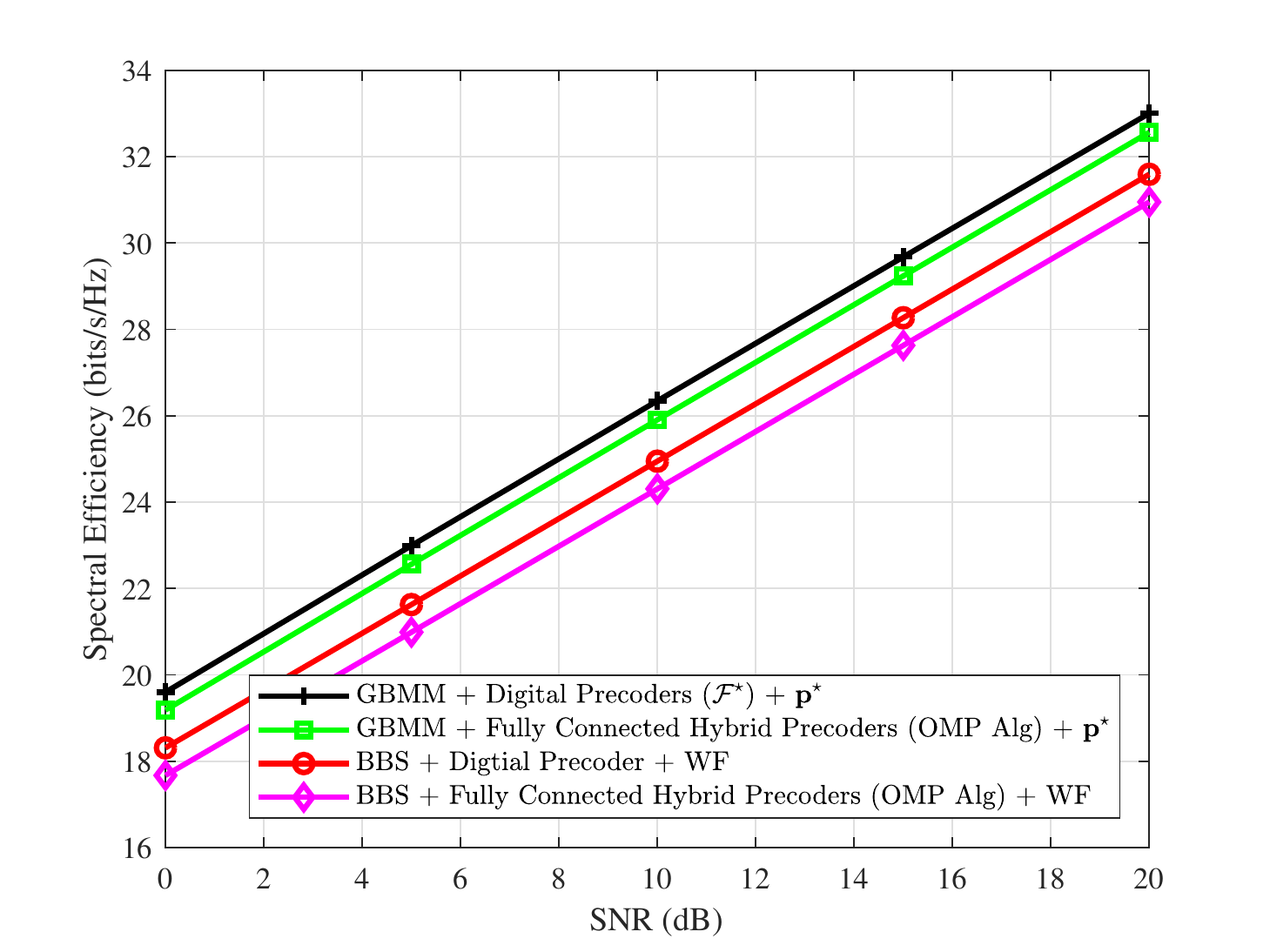}\\
 \caption{The SE of GBMM mmWave MIMO-OFDM where $N_t=100$, $N_r=36$, $N_{RF}^t=N_s=2$ and $K=128$.}
  \label{result4}
\end{figure}

Lastly, we investigate the extension to mmWave MIMO-OFDM systems as illustrated in Fig. \ref{result4}.  
Considering the \textbf{Algorithm 1} are of much higher complexity and will be prohibitive if the number of carriers is large, we use \textbf{Algorithm 2} to get the optimal fully-digital precoders and precoder activation  probability distribution.  Similarly,  OMP-Alg \cite{Ayach2014} is adopted to solve the problem (\textbf{P5}) for full-connected hybrid precoder designs, and all simulation results are averaged over $100$ channel realizations. In the simulations, $N_{RF}^{t}=N_s=2$ and the number of carriers $K=128$. BBS with digital precoder and that with fully-connected hybrid precoders  for each carrier are included as benchmarks. For BBS, water-filling power allocation is adopted, and we note that the water-filling power allocation approaches equal power allocation in the depicted SNR regime.
Results show that the proposed GBMM scheme can be directly extended to broadband systems to improve SE.

\section{Conclusions}
In this paper, we proposed a GBMM transmission solution for mmWave MIMO communications based on the concept of SM. GBMM uses precoder design to offer beamforming gain and beamspace indices to carry information which improves SE. Both analytical and simulation results showed that GBMM  outperforms the well recognized ``best'' BBS solution in mmWave MIMO systems.  Moreover, simulation results also demonstrated that compared with BBS and GBMM,  conventional SM techniques with equal precoder activation probability is the worst. Based on the derived theoretical SE, a gradient ascent algorithm was developed. With the proposed algorithm, the optimal precoders, precoder activation probability distribution and  power allocation for each data stream can be found. Simulation results validated its effectiveness and efficiency. Moreover,  the closed-form solutions of the optimal precoders, precoder activation probability distribution and  power allocation were derived in the high SNR regime, which can greatly reduce the system implementation complexity. Moreover, we discussed the impact of hybrid transceiver structure on SE and the extension to mmWave MIMO-OFDM systems. Results show that the proposed GBMM with existing hybrid designs can achieve better performance than that of BBS and can also be extended to broadband systems to improve SE.


\begin{thebibliography}{10}
\providecommand{\url}[1]{#1}
\csname url@samestyle\endcsname
\providecommand{\newblock}{\relax}
\providecommand{\bibinfo}[2]{#2}
\providecommand{\BIBentrySTDinterwordspacing}{\spaceskip=0pt\relax}
\providecommand{\BIBentryALTinterwordstretchfactor}{4}
\providecommand{\BIBentryALTinterwordspacing}{\spaceskip=\fontdimen2\font plus
\BIBentryALTinterwordstretchfactor\fontdimen3\font minus
  \fontdimen4\font\relax}
\providecommand{\BIBforeignlanguage}[2]{{%
\expandafter\ifx\csname l@#1\endcsname\relax
\typeout{** WARNING: IEEEtran.bst: No hyphenation pattern has been}%
\typeout{** loaded for the language `#1'. Using the pattern for}%
\typeout{** the default language instead.}%
\else
\language=\csname l@#1\endcsname
\fi
#2}}
\providecommand{\BIBdecl}{\relax}
\BIBdecl

\bibitem{Bai2014}
T.~Bai, A.~Alkhateeb, and R.~W. Heath, ``Coverage and capacity of
  millimeter-wave cellular networks,'' \emph{{IEEE} Commun. Mag.}, vol.~52,
  no.~9, pp. 70--77, Sep. 2014.

\bibitem{Yu2016}
X.~Yu, J.-C. Shen, J.~Zhang, and K.~B. Letaief, ``Alternating minimization
  algorithms for hybrid precoding in millimeter wave mimo systems,'' \emph{IEEE
  J. Sel. Topics Signal Process.}, vol.~10, no.~3, pp. 485--500, Feb. 2016.

\bibitem{Ayach2014}
O.~E. Ayach, S.~Rajagopal, S.~Abu-Surra, Z.~Pi, and R.~W. Heath, ``Spatially
  sparse precoding in millimeter wave {MIMO} systems,'' \emph{{IEEE} Trans.
  Wireless Commun.}, vol.~13, no.~3, pp. 1499--1513, Mar. 2014.

\bibitem{Chen2015}
C.~Chen, ``An iterative hybrid transceiver design algorithm for millimeter wave
  {MIMO} systems,'' \emph{IEEE Wireless Communications Letters}, vol.~4, no.~3,
  pp. 285--288, Jun. 2015.

\bibitem{Chen2017}
C.~Chen, C.~Tsai, Y.~Liu, W.~Hung, and A.~Wu, ``Compressive sensing ({CS})
  assisted low-complexity beamspace hybrid precoding for millimeter-wave {MIMO}
  systems,'' \emph{{IEEE} Trans. Signal Process.}, vol.~65, no.~6, pp.
  1412--1424, Mar. 2017.

\bibitem{Chen2018}
Y.~Chen, D.~Chen, Y.~Tian, and T.~Jiang, ``Low complexity hybrid precoding and
  diversity combining based on spatial lobes division for millimeter wave mimo
  systems,'' \emph{arXiv preprint arXiv:1803.00322}, 2018.

\bibitem{Dai2015}
L.~Dai, X.~Gao, J.~Quan, S.~Han, and C.~I, ``Near-optimal hybrid analog and
  digital precoding for downlink mmwave massive {MIMO} systems,'' in
  \emph{Proc. IEEE ICC 2015}, London, UK, Jun. 2015, pp. 1334--1339.

\bibitem{Han2015}
S.~Han, C.~I, Z.~Xu, and C.~Rowell, ``Large-scale antenna systems with hybrid
  analog and digital beamforming for millimeter wave {5G},'' \emph{{IEEE}
  Commun. Mag.}, vol.~53, no.~1, pp. 186--194, Jan. 2015.

\bibitem{He2016}
S.~He, C.~Qi, Y.~Wu, and Y.~Huang, ``Energy-efficient transceiver design for
  hybrid sub-array architecture {MIMO} systems,'' \emph{IEEE Access}, vol.~4,
  pp. 9895--9905, 2016.

\bibitem{Gao2016}
X.~Gao, L.~Dai, S.~Han, C.~I, and R.~W. Heath, ``Energy-efficient hybrid analog
  and digital precoding for mmwave mimo systems with large antenna arrays,''
  \emph{IEEE J. Sel. Areas Commun.}, vol.~34, no.~4, pp. 998--1009, Apr. 2016.

\bibitem{Li2017}
A.~Li and C.~Masouros, ``Hybrid analog-digital millimeter-wave mu-mimo
  transmission with virtual path selection,'' \emph{IEEE Commun. Lett.},
  vol.~21, no.~2, pp. 438--441, Feb. 2017.

\bibitem{Liang2014}
L.~Liang, W.~Xu, and X.~Dong, ``Low-complexity hybrid precoding in massive
  multiuser mimo systems,'' \emph{{IEEE} Commun. Lett.}, vol.~3, no.~6, pp.
  653--656, Dec. 2014.

\bibitem{Molu2018}
M.~M. Molu, P.~Xiao, M.~Khalily, K.~Cumanan, L.~Zhang, and R.~Tafazolli,
  ``Low-complexity and robust hybrid beamforming design for multi-antenna
  communication systems,'' \emph{IEEE Trans. Wireless Commun.}, vol.~17, no.~3,
  pp. 1445--1459, Mar. 2018.

\bibitem{Rusu2016}
C.~Rusu, R.~M{��}ndez-Rial, N.~Gonz��lez-Prelcic, and R.~W. Heath, ``Low
  complexity hybrid precoding strategies for millimeter wave communication
  systems,'' \emph{{IEEE} Trans. Wireless Commun.}, vol.~15, no.~12, pp.
  8380--8393, Dec. 2016.

\bibitem{Park2017}
S.~Park, A.~Alkhateeb, and R.~W. Heath, ``Dynamic subarrays for hybrid
  precoding in wideband mmwave {MIMO} systems,'' \emph{{IEEE} Trans. Wireless
  Commun.}, vol.~16, no.~5, pp. 2907--2920, May 2017.

\bibitem{Liao2017}
A.~Liao, Z.~Gao, Y.~Wu, H.~Wang, and M.~Alouini, ``{2D} unitary {ESPRIT} based
  super-resolution channel estimation for millimeter-wave massive {MIMO} with
  hybrid precoding,'' \emph{IEEE Access}, vol.~5, pp. 24\,747--24\,757, 2017.

\bibitem{Mao2018}
J.~Mao, Z.~Gao, Y.~Wu, and M.~Alouini, ``Over-sampling codebook-based hybrid
  minimum sum-mean-square-error precoding for millimeter-wave {3D-MIMO},''
  \emph{IEEE Wireless Communications Letters}, pp. 1--1, 2018.

\bibitem{Mesleh2008}
R.~Mesleh, H.~Haas, S.~Sinanovic, C.~W. Ahn, and S.~Yun, ``Spatial
  modulation,'' \emph{{IEEE} Trans. Veh. Technol.}, vol.~57, no.~4, p. 2228,
  Jul. 2008.

\bibitem{Yang2008}
Y.~Yang and B.~Jiao, ``Information-guided channel-hopping for high data rate
  wireless communication,'' \emph{{IEEE} Commun. Lett.}, vol.~12, no.~4, pp.
  225--227, Apr. 2008.

\bibitem{Renzo2014}
M.~D. Renzo, H.~Haas, A.~Ghrayeb, S.~Sugiura, and L.~Hanzo, ``Spatial
  modulation for generalized {MIMO}: Challenges, opportunities, and
  implementation,'' \emph{Proc. {IEEE}}, vol. 102, no.~1, pp. 56--103, Jan.
  2014.

\bibitem{Ishikawa2018}
N.~Ishikawa, S.~Sugiura, and L.~Hanzo, ``50 years of permutation, spatial and
  index modulation: From classic {RF} to visible light communications and data
  storage,'' \emph{{IEEE} Commun. Surveys Tuts.}, vol.~20, no.~3, pp.
  1905--1938, 3rd Quarter 2018.

\bibitem{Liu2015}
P.~Liu and A.~Springer, ``Space shift keying for {LOS} communication at mmwave
  frequencies,'' \emph{IEEE Wireless Commun. Lett.}, vol.~4, no.~2, pp.
  121--124, Apr. 2015.

\bibitem{Yang2017}
P.~Yang, Y.~Xiao, Y.~L. Guan, Z.~Liu, S.~Li, and W.~Xiang, ``Adaptive {SM-MIMO}
  for mmwave communications with reduced {RF} chains,'' \emph{IEEE J. Sel.
  Areas Commun.}, vol.~35, no.~7, pp. 1472--1485, Jul. 2017.

\bibitem{Mesleh2015}
R.~Mesleh, S.~S. Ikki, and H.~M. Aggoune, ``Quadrature spatial modulation,''
  \emph{IEEE Trans. Veh.Technol.}, vol.~64, no.~6, pp. 2738--2742, Jun. 2015.

\bibitem{Younis2017}
A.~Younis, N.~Abuzgaia, R.~Mesleh, and H.~Haas, ``Quadrature spatial modulation
  for {5G} outdoor millimeter�cwave communications: Capacity analysis,''
  \emph{IEEE Trans. Wireless Commun.}, vol.~16, no.~5, pp. 2882--2890, May
  2017.

\bibitem{Ishikawa2017}
N.~Ishikawa, R.~Rajashekar, S.~Sugiura, and L.~Hanzo,
  ``Generalized-spatial-modulation-based reduced-{RF}-chain millimeter-wave
  communications,'' \emph{IEEE Trans.Veh. Technol.}, vol.~66, no.~1, pp.
  879--883, Jan. 2017.

\bibitem{Liu2016}
C.~Liu, M.~Ma, Y.~Yang, and B.~Jiao, ``Optimal spatial-domain design for
  spatial modulation capacity maximization,'' \emph{{IEEE} Commun. Lett.},
  vol.~20, no.~6, pp. 1092--1095, Jun. 2016.

\bibitem{Liu2018}
P.~Liu, J.~Blumenstein, N.~S. Perovi, M.~D. Renzo, and A.~Springer,
  ``Performance of generalized spatial modulation {MIMO} over measured {60GHz}
  indoor channels,'' \emph{IEEE Trans. Commun.}, vol.~66, no.~1, pp. 133--148,
  Jan. 2018.

\bibitem{He2017}
L.~He, J.~Wang, and J.~Song, ``Spectral-efficient analog precoding for
  generalized spatial modulation aided mmwave {MIMO},'' \emph{IEEE Trans.Veh.
  Technolo.}, vol.~66, no.~10, pp. 9598--9602, Oct. 2017.

\bibitem{He2018}
------, ``Spatial modulation for more spatial multiplexing: {RF}-chain-limited
  generalized spatial modulation aided mm-wave {MIMO} with hybrid precoding,''
  \emph{{IEEE} Trans. Commun.}, vol.~66, no.~3, pp. 986--998, Mar. 2018.

\bibitem{Perovic2017}
N.~S. Perovi, P.~Liu, M.~D. Renzo, and A.~Springer, ``Receive spatial
  modulation for {LOS} mmwave communications based on {TX} beamforming,''
  \emph{{IEEE} Commun. Lett.}, vol.~21, no.~4, pp. 921--924, Apr. 2017.

\bibitem{Lee2017}
M.~Lee and W.~Chung, ``Adaptive multimode hybrid precoding for single-{RF}
  virtual space modulation with analog phase shift network in {MIMO} systems,''
  \emph{IEEE Trans. Wireless Commun.}, vol.~16, no.~4, pp. 2139--2152, Apr.
  2017.

\bibitem{Wang2018}
W.~Wang and W.~Zhang, ``Transmit signal designs for spatial modulation with
  analog phase shifters,'' \emph{IEEE Trans. Wireless Commun.}, vol.~17, no.~5,
  pp. 3059--3070, May 2018.

\bibitem{Ding2017}
Y.~Ding, K.~J. Kim, T.~Koike-Akino, M.~Pajovic, P.~Wang, and P.~Orlik,
  ``Spatial scattering modulation for uplink millimeter-wave systems,''
  \emph{{IEEE} Commun. Lett.}, vol.~21, no.~7, pp. 1493--1496, Jul. 2017.

\bibitem{Guo2019}
S.~Guo, P.~Zhang, P.~Zhao, L.~Wang, H.~Zhang, and M.-S. Alouini, ``Generalized
  beamspace modulation using multiplexing for mm-wave {MIMO},'' in
  \emph{submitted to IEEE ICC}, 2019.

\bibitem{Ibrahim2016}
A.~A.~I. Ibrahim, T.~Kim, and D.~J. Love, ``On the achievable rate of
  generalized spatial modulation using multiplexing under a {Gaussian} mixture
  model,'' \emph{{IEEE} Trans. Commun.}, vol.~64, no.~4, pp. 1588--1599, Apr.
  2016.

\bibitem{Boyd2004}
S.~Boyd and L.~Vandenberghe, \emph{Convex Optimization}.\hskip 1em plus 0.5em
  minus 0.4em\relax Cambridge University Press, 2004.

\bibitem{Goldsmith2003}
A.~Goldsmith, S.~A. Jafar, N.~Jindal, and S.~Vishwanath, ``Capacity limits of
  {MIMO} channels,'' \emph{{IEEE} J. Sel. Areas Commun.}, vol.~21, no.~5, pp.
  684--702, Jun. 2003.

\bibitem{Khakurel2014}
S.~Khakurel, C.~Leung, and T.~Le-Ngoc, ``A generalized water-filling algorithm
  with linear complexity and finite convergence time,'' \emph{IEEE Wireless
  Communications Letters}, vol.~3, no.~2, pp. 225--228, Apr. 2014.

\bibitem{Romanofsky2007}
R.~R. Romanofsky, \emph{Array phase shifters: Theory and technology}.\hskip 1em
  plus 0.5em minus 0.4em\relax Antenna Engineering Handbook, 4th ed, New York,
  NY, USA: McGraw-Hill, 2007., 2007.

\bibitem{Co2006}
N.~Co, ``A low cost analog phase shifter product family for military,
  commercial and public safety applications,'' \emph{Microw. J.}, vol.~49,
  no.~3, pp. 152--156, Mar. 2006.

\bibitem{Guo2016}
S.~Guo, H.~Zhang, S.~Jin, and P.~Zhang, ``Spatial modulation via {3-D}
  mapping,'' \emph{{IEEE} Commun. Lett.}, vol.~20, no.~6, pp. 1096--1099, Jun.
  2016.

\bibitem{Guo2017}
S.~Guo, H.~Zhang, P.~Zhang, D.~Wu, and D.~Yuan, ``Generalized {3-D}
  constellation design for spatial modulation,'' \emph{{IEEE} Trans. Commun.},
  vol.~65, no.~8, pp. 3316--3327, Aug. 2017.

\bibitem{Wang2016}
W.~Wang and W.~Zhang, ``Adaptive spatial modulation using {Huffman} coding,''
  in \emph{IEEE GLOBECOM}, Washington, DC USA, Dec. 2016, pp. 1--6.

\bibitem{Wang2017}
------, ``Huffman coding-based adaptive spatial modulation,'' \emph{{IEEE}
  Trans. Wireless Commun.}, vol.~16, no.~8, pp. 5090--5101, Aug. 2017.

\bibitem{Sohrabi2016}
F.~Sohrabi and W.~Yu, ``Hybrid analog and digital beamforming for {OFDM}-based
  large-scale {MIMO} systems,'' in \emph{IEEE SPAWC}, Edinburgh, UK, Jul. 2016,
  pp. 1--6.

\end{thebibliography}
\end{document}